\documentclass[11pt]{article}

\usepackage{amssymb,amsmath,amsfonts,amsthm,lscape}

\usepackage{graphicx}
 \usepackage{doi}

\newcommand{\be}{\begin{equation}}
\newcommand{\ee}{\end{equation}}
\newcommand{\bea}{\begin{eqnarray}}
\newcommand{\eea}{\end{eqnarray}}

\newcommand{\kk}{\kappa}
\newcommand{\bfI}{{\mathbb I}_\kappa}

\newcommand{\Sk}{{\rm\ \!S}}            
\newcommand{\Ck}{{\rm\ \!C}}           
 \newcommand{\Tk}{{\rm\ \!T}}           
  \newcommand{\pp}{\pi}

  \newcommand{\mm}{\mathcal{E}_\kk}
\newcommand{\xx}{X}

\newcommand{\yy}{Y}

\newcommand{\ang}{\mathcal{J}}

 \def\>#1{{\mathbf#1}}                 
\def\pois#1#2{\left\{ #1,#2 \right\}}  

 \def\k{{\kappa}}

 \newcommand{\bq}{\mathbf{q}}
\newcommand{\bp}{\mathbf{p}}

\newcommand{\tq}{ {q}}

   \newcommand{\del}{\delta}
  \newcommand{\Om}{\Omega}

\begin{document}

\thispagestyle{empty}

\

 \vskip1cm

 \begin{center}

  {\LARGE{\bf {Curvature as an integrable deformation}}}

\medskip
\medskip
\medskip
\medskip

 {\sc  Angel Ballesteros,  Alfonso Blasco  and Francisco J. Herranz}  
\medskip
\medskip

  {Departamento de F{\'{\i}}sica, Universidad de Burgos, 
09001 Burgos, Spain }
\medskip

 {E-mails:  {\tt    angelb@ubu.es,  ablasco@ubu.es, fjherranz@ubu.es}}

\end{center}

\medskip
\medskip

\begin{abstract}
\noindent
The generalization of (super)integrable Euclidean classical Hamiltonian systems to the two-dimensional sphere and the hyperbolic space by preserving their (super)integrability properties is reviewed. The constant Gaussian curvature of the underlying spaces is introduced as an explicit deformation parameter, thus allowing the construction of new integrable Hamiltonians in a unified geometric setting  in which the Euclidean systems are obtained in the vanishing curvature limit. In particular, the constant curvature analogue of the generic anisotropic oscillator Hamiltonian is presented, and its superintegrability for commensurate frequencies is shown. As a second example, an integrable version of the H\'enon--Heiles system on the sphere and the hyperbolic plane is introduced. Projective Beltrami coordinates are shown to be helpful in this construction, and further applications of this approach are sketched.\footnote{To appear in ``Integrability, Supersymmetry and Coherent States", 
 A volume in honour of Professor V\'eronique Hussin. 
 S. Kuru, J. Negro and L.M. Nieto (Eds.),  Special volume of the CRM Series in Mathematical Physics (Berlin: Springer, 2019)}
\end{abstract}

\medskip\medskip\medskip\medskip\medskip
%%%%%%%%%%%%%%%%%%%%%%%%%%%%%%%%%%%%%%%%%%%%%%%


  \medskip


  \medskip


\newpage

%%%%%%%%%%%%%%%%%%%%%%%%%%%%%%%%%%%%%%

%%%%%%%%%%%%%%%%%%%%%%%%%%%%%%%%%%%%%%%%%%%%%%%

\tableofcontents

 \contentsline {section}{Acknowledgements}{27}{section*}

   \contentsline {section}{References}{28}{section*}

%%%%%%%%%%%%%%%%%%%%%%%%%%%%%%%%%%%%%%%%%%%%%%%

\section{Introduction}
\label{intro}

The aim of this contribution is to review some new  recent results related to a seemingly elementary issue in the theory of finite-dimensional integrable systems~\cite{Perelomov,Goriely,Vozmischeva,BoPu,MiPWJPa13}, whose solution presents quite a number of interesting features. The problem can explicitly be  stated as follows.

Let us consider a certain Liouville integrable natural Hamiltonian system for a particle with unit mass moving on the two-dimensional (2D)  Euclidean space endowed with the standard bracket $\{q_{i},p_{j}\}=\delta_{ij}$ in terms of canonical coordinates and momenta, namely
\be
\mathcal{H}=\mathcal{T} + \mathcal{V}=\dfrac{1}{2}(p_{1}^{2}+p_{2}^{2}) + \mathcal{V}(q_1,q_2),
\label{Heucl}
\ee
where $\mathcal{T} $ is the kinetic energy and $\mathcal{V}$ is the potential. 
The Liouville integrability of this system will be provided by a constant of the motion given by a globally defined function $\mathcal{I}(p_1,p_2,q_1,q_2)$ such that $\pois{\mathcal{H}}{\mathcal{I}}=0$. 

The proposed problem consists in
finding a one-parameter integrable deformation of $\mathcal{H}$ of the form
\be
\mathcal{H}_\kappa=\mathcal{T}_\kappa(p_1,p_2,q_1,q_2) + \mathcal{V}_\kappa(q_1,q_2),
\qquad
\kappa\in \mathbb{R},
\nonumber
\ee
with integral of the motion given by the smooth and globally defined function $\mathcal{I}_\kappa(p_1,p_2,q_1,q_2)$ (therefore $\pois{\mathcal{H}_\kappa}{\mathcal{I}_\kappa}=0$), and such that the following two conditions hold:

\begin{enumerate}

\item The smooth function $\mathcal{T}_\kappa$ is the kinetic energy of a particle on a 2D space whose constant curvature is given by the parameter $\kappa$, {\em i.e.} the 2D sphere \textbf{S}$^{2}$ will arise in the case $\kappa>0$ and the hyperbolic plane  $\textbf{H}^{2}$ when $\kappa<0$.\\

\item The Euclidean system $\mathcal{H}$ given by~\eqref{Heucl} has to be smoothly recovered in the zero-curvature limit $\kappa\to 0$, namely
\be
\mathcal{H}=\lim_{\kappa\to 0}{\mathcal{H}_\kappa},
\qquad
\mathcal{I}=\lim_{\kappa\to 0}{\mathcal{I}_\kappa}.
\nonumber
\ee

\end{enumerate}

If these two conditions are fulfilled, we will say that $\mathcal{H}_\kappa$ is an {\em integrable curved version of $\mathcal{H}$} on the sphere and the hyperbolic space. We stress that within this framework the Gaussian curvature $\kappa$  of the space  enters as a  deformation parameter, and the curved system $\mathcal{H}_\kappa$ can be thought of as smooth integrable perturbation of the flat one $\mathcal{H}$ in terms of the curvature parameter. Therefore, integrable Hamiltonian systems on \textbf{S}$^{2}$ ($\kappa >0$),  \textbf{H}$^{2}$ ($\kappa <0$) and \textbf{E}$^{2}$ ($\kappa =0$) will be simultaneously constructed and analysed.

Moreover, it could happen that the initial Hamiltonian $\mathcal{H}$  is not only integrable but superintegrable, {\em i.e.}~another globally defined and functionally independent integral of the motion $\mathcal{K}(p_1,p_2,q_1,q_2)$ does exist such that
\be
\pois{\mathcal{H}}{\mathcal{I}}=\pois{\mathcal{H}}{\mathcal{K}}=0,
\qquad
\pois{\mathcal{I}}{\mathcal{K}}\neq 0.
\nonumber
\ee
In that case we could further impose the existence of the curved (and functionally independent) analogue $\mathcal{K}_\kappa$ of the second integral such that
\be
\mathcal{K}=\lim_{\kappa\to 0}{\mathcal{K}_\kappa} .
\nonumber
\ee If we succeed in finding such second integral fulfilling
\be
\pois{\mathcal{H_\kappa}}{\mathcal{I_\kappa}}=\pois{\mathcal{H_\kappa}}{\mathcal{K_\kappa}}=0,
\qquad
\pois{\mathcal{I_\kappa}}{\mathcal{K_\kappa}}\neq 0,
\nonumber
\ee
we will say that we have obtained a {\em superintegrable curved generalization} of the Euclidean superintegrable Hamiltonian $\mathcal{H}$.

The explicit curvature-dependent description of \textbf{S}$^{2}$ and \textbf{H}$^{2}$ is well-known in the literature and can be found, for instance, in~\cite{CK2,RS,trigo,conf,ran,ran1,BaHeSantS03,CRMVulpi,kiev,LetterBH,Santander6,CRS07JPa,CRS08,Kepler,DiacuJDE,Diacu1,Diacu2,DiacuM,GonKas14AnnPhys,Ra14JPaTTWk,Mfran15PLa,Ra15Jmp,Chanu} (see also references therein) where it has been mainly considered in the classification and description of superintegrable systems on these two spaces. In this contribution we will present several recent works in which this geometric framework has been applied for non-superintegrable systems where the lack of additional symmetries forces to make use of a purely integrable perturbation approach. Moreover, this perturbative viewpoint shows that the uniqueness of this construction is not guaranteed, since in general different $\mathcal{V}_\kappa$ integrable potentials (and their associated $\mathcal{I}_\kappa$ integrals) having the same $\kappa\to 0$ limit could exist and be found. As an outstanding example of this plurality, we will present the construction of different integrable curved analogues  on \textbf{S}$^{2}$ ($\kappa>0$) and \textbf{H}$^{2}$ ($\kappa<0$) of some anisotropic oscillators. 

The second novel technical aspect to be emphasized in the results here presented is that in some cases projective coordinates turn out to be helpful in order to construct the (super)integrable deformations $\mathcal{H}_\kappa$, since when these coordinates are considered on \textbf{S}$^{2}$ and \textbf{H}$^{2}$ then the curved kinetic energy $\mathcal{T}_\kappa$ is expressed as a polynomial in the canonical variables describing the projective phase space. Therefore, some of the examples here presented can be thought of as instances of integrable projective dynamics, in the sense of~\cite{Albouy2013,Albouy2015}.

The structure of the paper is the following. In the next Section we review the description of the geodesic dynamics on the sphere and the hyperboloid by making use of the above mentioned curvature-dependent formalism. In particular, ambient space coordinates as well as geodesic parallel and geodesic polar coordinates for  \textbf{S}$^{2}$ and \textbf{H}$^{2}$ will   be introduced. In Section 3 the projective dynamics on the sphere and the hyperboloid in terms of Beltrami coordinates will also be  summarized, thus providing a complete set of geometric possibilities for the description of dynamical systems on these curved spaces. In Section 4 we recall the (super)integrability properties of the 2D anisotropic  oscillator with   arbitrary frequencies and   also with commensurate ones, and in Section 5 the explicit construction of the $\mathcal{H_\kappa}$ Hamiltonian defining its curved analogue will be presented. Section 6 will be devoted to recall the three integrable versions of the well-known (non-integrable) H\'enon--Heiles Hamiltonian. In Section 7 the construction of the curved version on \textbf{S}$^{2}$ and \textbf{H}$^{2}$ of an integrable H\'enon--Heiles system related to the KdV hierarchy will be constructed, thus exemplifying the usefulness of the approach here presented for the obtention of new integrable systems on curved spaces. Furthermore, the full Ramani--Dorizzi--Grammaticos    series of integrable polynomial potentials will also be  generalized to the curved case. Finally, a Section including some remarks and open problems under investigation closes the paper.

%%%%%%%%%%%%%%%%%%%%%%%%%%%%%%%%%%%%%%%%%%%%%%%

\section{Geodesic dynamics on   the sphere and the hyperboloid}
\label{sec2}
 
  Let us consider the one-parametric family of 3D real Lie algebras $\mathfrak{so}_\kk(3)={\rm span}\{J_{01}, J_{02}, J_{12}\}$
    with commutation relations given by (in the sequel we follow the curvature-dependent formalism as presented in~\cite{BaHeMu13,BaBlHeMu14}):
\begin{equation}
  [J_{12},J_{01}]=J_{02},\qquad [J_{12},J_{02}]=-J_{01},\qquad [J_{01},J_{02}]=\kk J_{12}  , \label{ca}
 \end{equation}
where $\kk$ is a real parameter.    The Casimir invariant, coming from the  Killing--Cartan form, reads
\be
  {\cal C}=J_{01}^2+J_{02}^2+\kk J_{12}^2.
 \label{cb}
 \ee
The family $\mathfrak{so}_{\kk}(3)$ comprises    three specific Lie algebras:   $\mathfrak{so}(3)$ for $\kk>0$,  $\mathfrak{so}(2,1)\simeq \mathfrak{sl}_2(\mathbb R)$ for $\kk<0$, and  $\mathfrak{iso}(2)\equiv \mathfrak{e}(2) = \mathfrak{so}(2) \oplus_S \mathbb{R}^2$ for $\kk=0$.  Note that the  value of      $\kk$  can  be 
reduced to $\{+1 , 0 ,-1\}$ through a rescaling of the Lie algebra generators; therefore setting $\kk=0$ in (\ref{ca}) can be shown to be   equivalent to applying  an In\"on\"u--Wigner contraction~\cite{IW}.

The involutive
automorphism  defined by
$$
\Theta  (J_{01},J_{02},J_{12})=(-J_{01},-J_{02},J_{12}) ,
$$
generates a $\mathbb Z_2$-grading of   $\mathfrak{so}_\kk(3)$ in such a manner that   $\kk$ is  a
graded contraction parameter~\cite{Montigny}, and  $\Theta$  gives  rise to the following Cartan
decomposition of the Lie algebra:
$$
 \mathfrak{so}_\kk(3) ={\mathfrak{h}}  \oplus  {\mathfrak{p}} ,\qquad 
{\mathfrak{h}  }={\rm span}\{ J_{12} \} =\mathfrak{so}(2) ,\qquad
{\mathfrak{p}  }={\rm span}\{ J_{01},J_{02} \}  .
$$
We denote $\rm{SO}_\kk(3)$ and $H$ the Lie groups with Lie algebras $ \mathfrak{so}_\kk(3) $ and $\mathfrak{h}$, respectively, and 
we  consider the 2D symmetrical homogeneous
space defined by 
\be
\mathbf{S}^2_\kk = {\rm  SO}_{\kk}(3)/H ,\qquad H= {\rm  SO}(2) =\langle J_{12}\rangle .
\label{cc}
\ee
This coset space  has   constant Gaussian curvature equal to $\kk$ and  is endowed with a  metric having positive  definite signature. 
The generator  $J_{12}$ leaves a point $O$ invariant, the origin, so generating  
 rotations around $O$, while $J_{01}$ and $J_{02}$    generate translations which 
 move $O$ along  two basic orthogonal geodesics $l_1$ and $l_2$.

Therefore $\mathbf{S}^2_\kk $ (\ref{cc}) covers the three classical 2D Riemannian spaces of constant curvature:
  $$
\begin{array}{lll}
{\mathbf
S}^2_+:\ \mbox{ Sphere}&\qquad {\mathbf
S}^2_0:\ \mbox{Euclidean plane}&\qquad{\mathbf
S}^2_- :\ \mbox{Hyperbolic space}\\[2pt]
   {\mathbf S}^2={\rm SO}(3)/{\rm  SO}(2)&\qquad    {\mathbf E}^2={\rm  ISO}(2)/{\rm  SO}(2)&\qquad    {\mathbf H}^2= {\rm  SO}(2,1)/{\rm SO}(2)
\end{array}
$$
We recall that these three spaces (and their motion groups $ {\rm  SO}_{\kk}(3)$) are contained within the family of the so-called 2D orthogonal Cayley--Klein geometries~\cite{CK2,Yaglom,Groma}, which are parametrized in terms of two graded contraction parameters $\kk\equiv \kk_1$ and $\kk_2$~\cite{Montigny}.

In what follows we describe the metric structure  and the geodesic motion on the above spaces in terms of several sets of coordinates that will be used throughout the paper. We stress that   all the resulting expressions will have always a  smooth and well-defined flat limit (contraction) $\kk\to 0$  reducing to the corresponding Euclidean ones.

%%%%%%%%%%%%%%%%%%%%%%%%%%%%%%%%%%%%%%%%%%%%%%%

\subsection{Ambient space  coordinates}
\label{sec21}

The  {\em vector representation} of  $\mathfrak{so}_\kk(3)$ is provided by the  following  faithful matrix representation   $\rho:\mathfrak{so}_{\kk}(3)  \rightarrow \rm{End} (\mathbb{R}^3)$~\cite{trigo,conf}
\be
\rho(J_{01})=\left(\begin{array}{ccc}
0&\, -\kk \, &0\cr
1&0 & 0 \cr
0& 0 &0 
\end{array}\right),
\ \ 
\rho(J_{02})=\left(\begin{array}{ccc}
0\ & 0 &\, -\kk  \cr
0\ & 0 &  0\cr
1\ & 0 & 0
\end{array}\right) ,\ \  
\rho(J_{12})=\left(\begin{array}{ccc}
0\ & 0 & 0 \cr
0\ & 0 & -1 \cr
0\ & 1 & 0
\end{array}\right),
\label{cd}
\ee
which satisfies
\begin{equation}
\rho(J_{ij})^T \bfI\,+\bfI \rho(J_{ij})=0,\qquad \bfI=\rm{diag}(1,\kk,\kk)  .
\label{ce}
\end{equation}
The matrix exponentiation of (\ref{cd}) leads to the following one-parametric subgroups of $\rm{SO}_\kk(3)$:
\begin{equation}
\begin{aligned}
{\rm e}^{\alpha \rho(J_{01})} & =\left(\begin{array}{ccc}
\Ck_{\kk}(\alpha)\, &-\kk\Sk_{\kk}(\alpha)\ &0 \\[1pt] 
\Sk_{\kk}(\alpha)\, &\Ck_{\kk}(\alpha)&0 \\[1pt] 
0&0&1
\end{array}\right)  , \quad  
{\rm e}^{\gamma \rho({J_{12} } )}=\left(\begin{array}{ccc}
1\ &0&0 \\[1pt] 
0\ &\cos\gamma&- \sin\gamma \\[1pt] 
0\ &\sin\gamma&\cos \gamma
\end{array}\right) , \\[2pt] 
{\rm e}^{\beta \rho(J_{02})}&  =\left(\begin{array}{ccc}
\Ck_{\kk}(\beta)\ &0\,&-\kk\Sk_{\kk}(\beta) \\[1pt] 
0&1\, &0 \\[1pt] 
\Sk_{\kk}(\beta)&0\,&\Ck_{\kk}(\beta)
\end{array}\right)  ,
\end{aligned}
\label{cf}
\end{equation}
where we have introduced the $\kk$-dependent cosine  and   sine  functions~\cite{CK2,trigo}
\begin{equation}
\Ck_{\kk}(x):=\sum_{l=0}^{\infty}(-\kk)^l\frac{x^{2l}} 
{(2l)!}=\left\{
\begin{array}{ll}
  \cos{\sqrt{\kk}\, x} &\quad  \kk>0 \\ 
\qquad 1  &\quad
  \kk=0 \\
\cosh{\sqrt{-\kk}\, x} &\quad   \kk<0 
\end{array}\right.  
\nonumber
\end{equation}
\begin{equation}
   \Sk{_\kk}(x) :=\sum_{l=0}^{\infty}(-\kk)^l\frac{x^{2l+1}}{ (2l+1)!}
= \left\{
\begin{array}{ll}
  \frac{1}{\sqrt{\kk}} \sin{\sqrt{\kk}\, x} &\quad  \kk>0 \\ 
\qquad x  &\quad
  \kk=0 \\ 
\frac{1}{\sqrt{-\kk}} \sinh{\sqrt{-\kk}\, x} &\quad  \kk<0 
\end{array}\right.  .
\nonumber
\end{equation}
The
$\kk$-tangent function  is defined as
\be
\Tk_\kk(x)  := \frac{\Sk_\kk(x)}  { \Ck_\kk(x)} \, .
\nonumber
\ee
These  curvature-dependent trigonometric functions coincide with  the  circular
and hyperbolic  ones for   $\kk=\pm 1$, while under the
contraction      $\kk=0$ they reduce    to the parabolic 
functions:  $\Ck_{0}(x)=1$ and 
$\Sk_{0}(x)=\Tk_{0}(x)=x$.  Some  trigonometric relations   read~\cite{trigo}
$$
 \Ck^2_\kk(x)+\kk\Sk^2_\kk(x)=1,  \ \   \Ck_\kk(2x)= \Ck^2_\kk(x)-\kk\Sk^2_\kk(x), \ \  \Sk_\kk(2x)= 2 \Sk_\kk(x) \Ck_\kk(x) 
 $$
 and their derivatives are given by~\cite{conf}
 $$
  \frac{ {\rm d}}{{\rm d} x}\Ck_\kk(x)=-\kk\Sk_\kk(x),\qquad        \frac{ {\rm d}}
{{\rm d} x}\Sk_\kk(x)= \Ck_\kk(x)  ,\qquad    
\frac{ {\rm d}}
{{\rm d} x}\Tk_\kk(x)=  \frac{1}{\Ck^2_\kk(x) } \,  .
$$

Therefore, under the matrix realization (\ref{cf}),  the Lie group ${\rm SO}_{\kk}(3)$ becomes a group of  isometries of  the bilinear form  $\bfI$ (\ref{ce}), 
$$
  g^T \bfI\, g=\bfI  ,\quad \forall g\in {\rm SO}_{\kk}(3)  ,
$$
 acting on a 3D linear ambient space $\mathbb{R}^3=(x_0,x_1,x_2)$ through matrix multiplication.
The  subgroup ${\rm e}^{\gamma \rho({J_{12} } )}$ (\ref{cf})      is the isotropy subgroup of the point
$O=(1,0,0)$, which  is     taken as the {\em origin} in the homogeneous space
$\mathbf S^2_\kk$ (\ref{cc}).   The orbit of  $O$ is contained in the ``$\kk$-sphere'' determined by $\bfI$  (\ref{ce}):
\begin{equation}
\Sigma_\kk\,  :\ x_0^2+\kk \bigl( x_1^2+
  x_2^2 \bigr)=1 .
\label{ci}
\end{equation}
The connected component of $\Sigma_\kk$  is identified with the space  ${\mathbf
S}^2_{\kk}$ and the action of  ${\rm SO}_{\kk}(3)$ is transitive on it.  The coordinates $(x_0,x_1,x_2)$,  satisfying  the constraint (\ref{ci})  are called {\em ambient space} or {\em Weierstrass coordinates}.
Notice that  for $\kk>0$ we recover the   sphere, if $\kk<0$ we find the two-sheeted hyperboloid, and in the flat case with $\kk=0$ we get 
 two Euclidean planes $x_0=\pm 1$ with Cartesian coordinates $(x_1,x_2)$.   Since  $O=(1,0,0)$,  we   identify the hyperbolic space $\mathbf H^2$  with the connected  component corresponding to the   sheet of the hyperboloid with $x_0\ge 1$,  and the Euclidean space $\mathbf E^2$   with the plane $x_0=+1$.

The metric on ${\mathbf
S}^2_\kk$
comes from    the flat ambient metric in $\mathbb R^{3}$ divided by the curvature $\kk$ and
restricted to $\Sigma_\kk$:
\begin{equation}
({\rm d} s)_\kk^2=\left.\frac {1}{\kk}
\bigl({\rm d} x_0^2+   \kk \left( {\rm d} x_1^2+   {\rm d} x_2^2\right)
\bigr)\right|_{\Sigma_\kk} \!\! =    \frac{\kk\left(x_1{\rm d} x_1 + x_2{\rm d} x_2 \right)^2}{1-  \kk \left(x_1^2+   x_2^2\right)}+  {\rm d} x_1^2+   {\rm d} x_2^2\, .
\label{cj}
\end{equation}
Isometry vector fields in   ambient coordinates for $\mathfrak{so}_{\kk }(3)$, fulfilling     (\ref{ca}),  are directly  obtained  from the vector representation (\ref{cd}):
\be
J_{01}=\kk \, x_1 \partial_0  -x_0 {\partial_1}   ,\qquad
J_{02}=\kk  \,x_2  {\partial_0}     -x_0 {\partial_2}  ,\qquad 
J_{12}=  x_2 {\partial_1}    -x_1 {\partial_2}   ,
\label{ck}
\ee
where $\partial_\mu=\partial/{\partial x_\mu}$ $(\mu=0,1,2)$.

Now we consider the    ambient momenta $\pp_\mu$ conjugate to $x_\mu$   fulfilling the canonical Poisson bracket
$
\{ x_\mu, \pp_\nu\}=\delta_{\mu\nu} 
$
  subjected to the constraint (\ref{ci}). The vector fields (\ref{ck})   give rise to a  symplectic  realization    of
$\mathfrak{so}_{\kk }(3)$ in terms of ambient variables  by setting
$\partial_\mu\to -\pp_\mu$:
\begin{equation}
J_{01}=x_0 \pp_1-\kk\, x_1 \pp_0 ,\qquad
J_{02}= x_0 \pp_2-\kk\, x_2 \pp_0,\qquad 
J_{12}=  x_1 \pp_2-x_2 \pp_1 ,
\label{cl}
\end{equation}
which close the Poisson brackets defining the Lie--Poisson algebra $\mathfrak{so}_{\kk }(3)$ 
\begin{equation}
  \{J_{12},J_{01}\}=J_{02},\qquad \{J_{12},J_{02}\}=-J_{01},\qquad \{J_{01},J_{02}\}=\kk J_{12}  .
\nonumber
 \end{equation}
 The metric (\ref{cj}) provides the free Lagrangian ${\cal L}_\kk$  with ambient velocities
$\dot x_\mu$ for a particle with unit mass, so determining  geodesic motion on ${\mathbf
S}^2_\kk$:
\be
{\cal L}_\kk=  \left. \frac 1{2\kk} \bigl( \dot x_0^2+   \kk \left(  \dot x_1^2+  \dot  x_2^2 \bigr)  \right)
\right|_{\Sigma_\kk} \!\! =    \frac{\kk\left(x_1\dot  x_1 + x_2\dot  x_2 \right)^2}{2\left(1-  \kk \left(x_1^2+   x_2^2\right)\right)}+  \frac 12 \left( \dot  x_1^2+   \dot  x_2^2\right) .
\label{cm}
\ee
Thus the corresponding   momenta $\pp_\mu=\partial {\cal L}_\kk/\partial \dot x_\mu$   read
 \begin{equation}
\pp_0=\dot x_0/\kk,\qquad \pp_1=\dot x_1,\qquad \pp_2= \dot x_2 .
\label{cn}
\end{equation}
The time derivative of the  constraint   (\ref{ci}) provides the relation
$$
\Sigma_\kk\,  :\ x_0\pp_0+x_1\pp_1+x_2\pp_2=0 .
$$
Finally, by introducing (\ref{cn}) in (\ref{cm}) we obtain that the kinetic energy ${\cal T}_\kk$ in ambient variables  is given by
\be
{\cal T}_\kk= \left.  \frac 1{2 }\left(\kk\,   \pp_0^2+       \pp_1^2+     \pp_2^2  \right)
 \right |_{\Sigma_\kk}  \!\! =    \frac{\kk\left(x_1   \pp_1 + x_2   \pp_2 \right)^2}{2\left(1-  \kk \left(x_1^2+   x_2^2\right)\right)}+  \frac 12 \bigl(  \pp_1^2+      \pp_2^2\bigl).
 \label{co}
\ee
Notice that the contraction $\kk=0$ is well-defined in the r.h.s.~of the equations (\ref{cj}), (\ref{cm})  and (\ref{co}) yielding the Euclidean expresions
$$
({\rm d} s)_0^2=  {\rm d} x_1^2+   {\rm d} x_2^2 ,\qquad {\cal L}_0=     \tfrac 12  ( \dot  x_1^2+   \dot  x_2^2 ) , \qquad {\cal T}_0=   \tfrac 12  (  \pp_1^2+      \pp_2^2 ).
$$

%%%%%%%%%%%%%%%%%%%%%%%%%%%%%%%%%%%%%%%%%%%%%%%

\subsection{Geodesic parallel and polar  coordinates}
\label{sec22}

  The ambient coordinates  (\ref{ci})  can also be parametrized in terms of two intrinsic  variables of geodesic type. For our purposes let us consider the so-called  {\em  geodesic parallel} $(x,y)$    and  {\em geodesic polar} $(r,\phi)$ coordinates of a point $Q=(x_0,x_1,x_2)\in\mathbf S^2_{\kk}$~\cite{RS,conf}, which  are defined through the following action of the  one-parametric subgroups (\ref{cf}) on the origin $O=(1,0,0)$:
 \begin{equation}
\begin{aligned}
(x_0,x_1,x_2)^T& = \exp(x\rho(J_{01})) \exp(y \rho(J_{02}))O^T\cr
&=\exp(\phi \rho(J_{12})) \exp(r \rho(J_{01}))O^T  ,
\end{aligned}
\nonumber
\end{equation}
which gives
\begin{equation}
\begin{aligned}
 x_0&=\Ck_{\kk}(x)\Ck_{\kk}(y) =\Ck_{\kk}(r)  , \cr
  x_1&=\Sk_{\kk }(x)\Ck_{\kk }(y)= \Sk_{\kk }(r)\cos \phi  , \cr
  x_2& =\Sk_{\kk }(y)=\Sk_{\kk }(r)\sin \phi .
\end{aligned}
\label{eb}
\end{equation}
In this construction, the variable    $r$ is the   distance   between the origin   $O$ and the point $Q$ measured along  the  geodesic $l$ that joins  both points, while   $\phi$ is the   angle  of $l$  with respect to  a base geodesic $l_1$ (associated with the translation generator $J_{01}$). 
Let  $Q_1$ be   the intersection point of $l_1$ with its orthogonal geodesic $l_2'$ through $Q$. Then    $x$  is the geodesic distance between $O$ and $Q_1$ measured  along $l_1$  and    $y$ is the  geodesic distance   between $Q_1$ and $Q$ measured  along $l_2'$.  
On $\mathbf E^2$   with $\kk=0$, the relations (\ref{eb}) lead to  $x_0=1$ and
    $(x_1,x_2)=(x,y)=(r\cos\phi, r\sin\phi)$ so reducing   to Cartesian  and   polar coordinates. 
    
\begin{figure}
\centerline{\includegraphics[width=12.8cm]{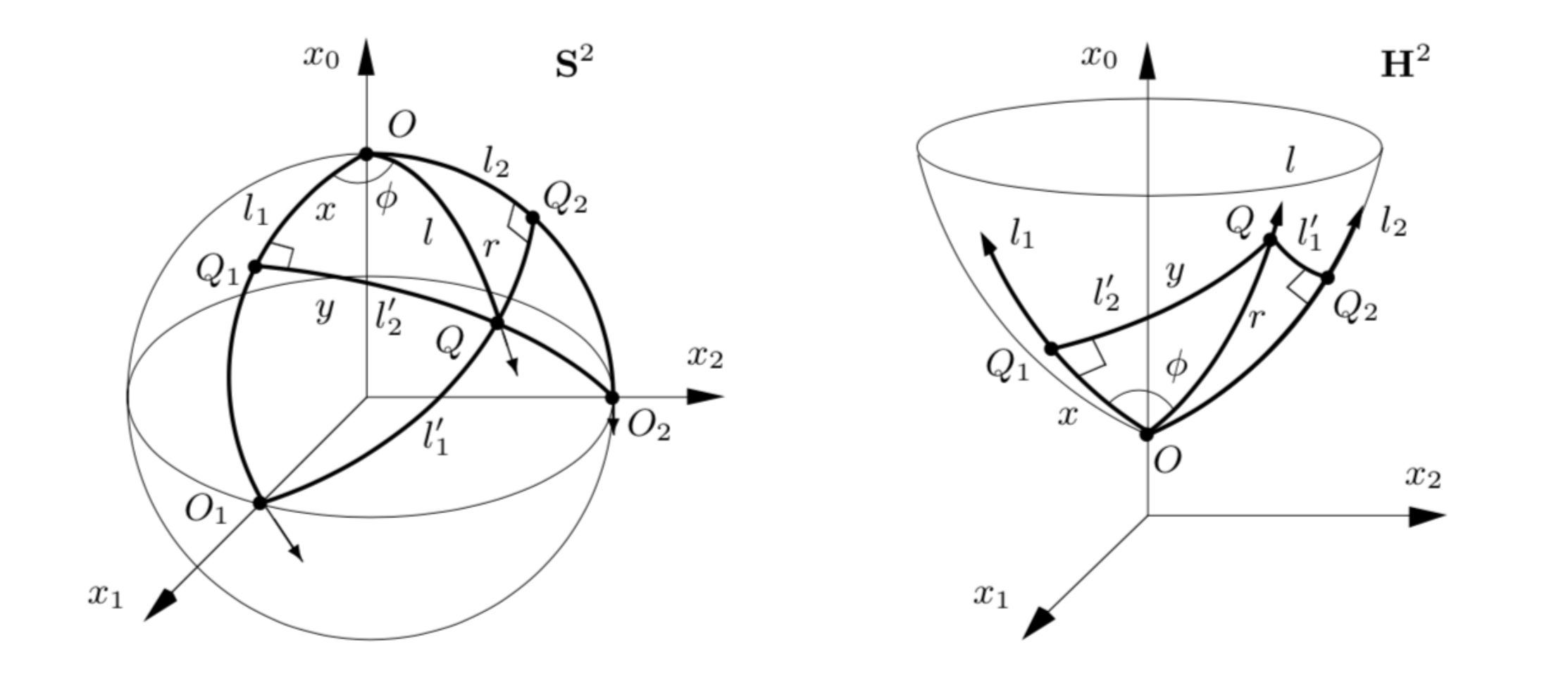}}
\caption{Ambient   $(x_0,x_1,x_2)$,  geodesic parallel $(x,y)$ and geodesic polar $(r,\phi)$ coordinates  of a point $Q$ on  the sphere  $ {\mathbf S}^2$ with $\kk=+1$ and on the hyperbolic space  $ {\mathbf H}^2$ with   $\kk=-1$ and $x_0\ge 1$. The origin of the space in ambient coordinates  is $O=(1,0,0)$. Note that  on $ {\mathbf S}^2$,   $O_1=(0,1,0)$ and $O_2=(0,0,1)$}
\label{figure2}       
\end{figure}

     These coordinates are shown Figure~\ref{figure2} for $\mathbf S^2$ and $\mathbf H^2$.  In these pictures, $l_2$ is the    base geodesic  orthogonal to $l_1$  through $O$, so related to      $J_{02}$, and  $Q_2$ is   the intersection point of $l_2$ with its orthogonal geodesic $l_1'$ through $Q$.

We substitute    (\ref{eb})   in the ambient  metric  (\ref{cj}) and in the free Lagrangian (\ref{cm}), finding that 
\begin{equation}
\begin{aligned}
({\rm d} s)_\kk^2 &=\Ck^2_{\kk}(y){\rm d} x^2 +  {\rm d} y^2 =      {\rm d} r^2+ \Sk^2_{\kk}(r)  {\rm d} \phi^2 ,\cr
 {\cal L}_\kk&= \tfrac12 \bigr(  \!\Ck^2_{\kk}(y) \dot x^2 +  \dot y^2  \bigl)=   \tfrac12 \bigr(    \dot r^2+ \Sk^2_{\kk}(r)  \dot \phi^2 \bigr).
\end{aligned}
\nonumber
\end{equation}

 Now, we   denote  $(p_x,p_y)$ and $(p_r,p_\phi)$ the conjugate momenta of  the coordinates $(x,y)$    and    $(r,\phi)$, respectively, 
 and  the free Hamiltonian (kinetic energy) turns out to be
\be
 {\cal T}_\kk=\frac 12\left(\frac{p_x^2}{ \Ck_{\kk}^2(y)} +p_y^2  \right)= \frac 12\left(p_r^2+  \frac{p_\phi^2}{ \Sk_{\kk}^2(r)}   \right).
\label{ed}
\ee

According to (\ref{eb}) and avoiding singularities in (\ref{ed}), we find that   the domain of the   geodesic coordinates on $\mathbf S^2$ and $\mathbf H^2$ reads (always $\phi\in[0,2\pi)$)
\bea
 {\mathbf S}^2\ (\kk>0) :&&      -\frac{\pi}{\sqrt{\kk}}< x\le \frac{\pi}{\sqrt{\kk}} ,\quad   -\frac{\pi}{2\sqrt{\kk}}< y< \frac{\pi}{2\sqrt{\kk}} ,  \quad  0< r< \frac{\pi}{\sqrt{\kk}}  .   \nonumber \\ 
 {\mathbf H}^2\ (\kk<0) :&&      -\infty< x< \infty ,\quad   -\infty< y< \infty,  \quad  0< r<\infty  .
 \label{eedd}
\eea

%%%%%%%%%%%%%%%%%%%%%%%%%%%%%%%%%%%%%%%%%%%%%%%

%%%%%%%%%%%%%%%%%%%%%%%%%%%%%%%%%%%%%%%%%%%%%%%

%%%%%%%%%%%%%%%%%%%%%%%%%%%%%%%%%%%%%%%%%%%%%%%

\section{Beltrami coordinates and projective dynamics}
\label{sec3}

 The quotients $(x_1/x_0,x_2/x_0)\equiv (q_1,q_2)$  of the ambient coordinates (\ref{ci}) are  just the   {\em Beltrami coordinates}  of projective geometry for the sphere and the hyperbolic plane. They are obtained by applying   
 the central stereographic projection with pole  
$(0,0,0)\in \mathbb R^{3}$ of a point $ Q=(x_0, x_1,x_2)$ onto the projective plane with $x_0=1$ and coordinates $(q_1,q_2)$:
$$
(x_0, x_1,x_2)\in\Sigma_\kk\ \to\  (0,0,0)+\mu\,
(1,q_1,q_2)\in \Sigma_\kk ,
$$
giving rise to the expressions
\begin{equation}
\begin{aligned}
x_0&=\mu=\frac{1}{\sqrt{1+ \kk (q_1^2+q_2^2)}},\qquad
x_i=\mu\, q_i=\frac{q_i}{\sqrt{1+ \kk (q_1^2+q_2^2)}},    \cr
 q_i&=\frac{x_i}{x_0},\qquad q_1^2+q_2^2= \frac{1-x_0^2}{\kk x_0^2} ,  \qquad  i=1,2.
\end{aligned}
\label{da}
\end{equation}
 Thus the   origin  $O=(1,0,0)\in \Sigma_\kk$  goes to the origin $(q_1,q_2)=(0,0)$ in the projective space ${\mathbf
S}^2_\kk$.

 The  domain of $(q_1,q_2)$   depends on the   value of the curvature $\kk$. We write $\kk$ in terms of the radius $R$ of the space as $\kk=\pm 1/R^2$ and we find that in the sphere ${\mathbf
S}^2 $ with $\kk=1/R^2>0$,    $q_i\in(-\infty,+\infty)$. The points in the equator in $ \Sigma_\kk$ with $x_0=0$ ($x_1^2+x_2^2=R^2$) go to infinity,  so that the projection (\ref{da}) is   well-defined for the   hemisphere  with $x_0>0$.
   In the hyperbolic or Lobachevski space  ${\mathbf
H}^2$  with $\kk=   -1/R^2<0$ and    $x_0\ge 1$ it is satisfied that
 $$
q_1^2+q_2^2= \frac{x_0^2-1}{|\kk|  x_0^2 } < R^2 ,
 $$
 which  is    the  {\em Poincar\'e disk} in Beltrami coordinates and
 $$
 q_i\in  \bigl(-1/\sqrt{|\kk|}, + 1/\sqrt{|\kk|} \bigr)=\left(-R, + R \right).
 $$  
 The points at the infinity in ${\mathbf
H}^2$ $(x_0\to\infty)$ are mapped onto  to the circle
$ q_1^2+q_2^2 = R^2  $.     Finally,  in the Euclidean plane ${\mathbf
E}^2$, with $\kk=0$  ($R\to \infty$), the Beltrami coordinates are just the Cartesian ones $x_i = q_i\in(-\infty,+\infty)$.

By introducing  (\ref{da}) in the ambient metric (\ref{cj}) and in the free Lagrangian (\ref{cm}) we obtain that
\be
({\rm d} s)_\kk^2= \frac{(1+\kk\, \>q^2) {\rm d} \>q^2 - \kk (\>q\cdot {\rm d }\>q)^2}{(1+\kk\, \>q^2)^2} ,\quad {\cal L}_\kk=  \frac{(1+\kk\, \>q^2) \dot{ \>q}^2 - \kk (\>q\cdot  \dot{ \>q})^2}{2(1+\kk\, \>q^2)^2} ,
\label{db}
\ee
where $\>q=(q_1,q_2)$ and  hereafter we shall use the following notation for any 2-vectors $\>a=(a_1,a_2)$ and $\>b=(b_1,b_2)$:
$$
\>a^2=a_1^2+a_2^2,\qquad \>a\cdot \>b= a_1 b_1 + a_2 b_2 .
$$
The  Beltrami momenta $\>p=(p_1,p_2)$ conjugate to the coordinates $\>q$, such that $\{ q_i,p_j\}=\delta_{ij}$,  come  from  $p_i=\partial {\cal L}_\kk/\partial \dot q_i$
\be
p_i= \frac{(1+\kk\,\>q^2)\dot q_i  - \kk (\>q \cdot { \dot {\>q} } )q_i} {(1+\kk\,\>q^2)^2} ,\qquad \dot q_i =   (1+\kk\,\>q^2)\bigl( p_i+ \kk (\>q \cdot {   {\>p} } )q_i  \bigr)
   .
\label{dcc}
\ee
And by inserting these expressions into $ {\cal L}_\kk$ (\ref{db}) we get 
 the free Hamiltonian  
\be
 {\cal T}_\kk=\tfrac 12  (1+\kk\,\>q^2)\bigl( \>p^2+ \kk (\>q \cdot {   {\>p} } )^2 \bigr) .
\label{dc}
\ee

By  introducing (\ref{da}) and   (\ref{dcc})  in  (\ref{cn})  we obtain   the
ambient momenta written in terms of the Beltrami variables,   $\pp_\mu(\>q,\>p)$, and from this result a  symplectic realization of the Lie--Poisson generators (\ref{cl}) in these variables is directly found. These expressions are displayed in Table~\ref{table1}. Notice  that  the kinetic energy (\ref{dc})  can also be recovered by computing the symplectic realization of the Casimir    (\ref{cb}) of $\mathfrak{so}_\kk(3)$ in   Beltrami variables as $ {\mathcal T}_\kk \equiv\frac 12 {\mathcal C}$.
Likewise the   ambient momenta $\pp_\mu$  and symplectic realization of the Lie--Poisson generators $J_{\mu\nu}$ can be computed in the  geodesic variables introduced in Section~\ref{sec22}, and these are also presented in Table~\ref{table1}.

We  recall that a similar procedure can be performed with Poincar\'e coordinates~\cite{Doubrovine} which come from 
 the stereographic projection with   pole  $(-1,0,0)$. The resulting expressions can be found in~\cite{BaBlHeMu14}.

   %%%%%%%%%%%%% Table 1 %%%%%%%%%%%%%%%%%%%%%%%%%%%%%%%

 \begin{landscape}

\begin{table}[t]
\caption{Expressions for the ambient variables $(x_\mu,\pp_\mu)$,   free Hamiltonian  $ {\cal T}_\kk$ and    symplectic realization of the Lie--Poisson generators $J_{\mu\nu}$  of $\mathfrak{so}_\kk(3)$ in terms of Beltrami, geodesic parallel and geodesic polar canonical variables. The specific expressions for ${\mathbf S}^2$,  ${\mathbf H}^2$ and ${\mathbf E}^2$  correspond to set  $\kk>0$, $\kk<0$ and $\kk=0$, respectively}
\label{table1}
 \begin{center}
\noindent
\begin{tabular}{l   l l l }
\hline
\\[-0.2cm]
\multicolumn{1}{l}{ }&
\multicolumn{1}{l}{Beltrami   $(\>q, \>p)$}&
\multicolumn{1}{l}{Geodesic parallel   $(x, y,p_x,p_y)\qquad$}
&\multicolumn{1}{l}{Geodesic polar   $(r,\phi,p_r,p_\phi)$}\\[0.2cm]
\hline
\\[-0.2cm] 
$x_0$&  $\displaystyle{\frac{1}{ ({1+ \kk \>q^2})^{1/2}  }      }  $   &  $ \Ck_{\kk}(x)\Ck_{\kk}(y) $&$ \Ck_{\kk}(r)$\\ [0.3cm]
$x_1$& $\displaystyle{ \frac{q_1}{ ({1+ \kk \>q^2})^{1/2}  }     }  $   &  $\Sk_{\kk }(x)\Ck_{\kk }(y) $   & $\Sk_{\kk}(r)\cos  \phi$\\ [0.3cm]
$x_2$&$ \displaystyle{\frac{q_2}{ ({1+ \kk \>q^2})^{1/2}  }      } $  &  $ \Sk_{\kk }(y)$  &$\Sk_{\kk}(r)\sin  \phi$    
\\[0.4cm]
\hline
\\[-0.2cm]
$\pp_0$&$\displaystyle{ -\sqrt{1+\kk\>q^2}\,(\>q\cdot \>p) } $ & $\displaystyle {-\frac{\Sk_{\kk}(x)}{\Ck_{\kk}(y)}\, p_x  - \Ck_{\kk}(x) \Sk_{\kk}(y) p_y}$ &$ - \Sk_{\kk}(r)\, p_r$  \\[0.3cm]
$\pp_1$&$\displaystyle{ \sqrt{1+\kk\>q^2}\, p_1 }  $ &$\displaystyle {\frac{\Ck_{\kk}(x)}{\Ck_{\kk}(y)}\, p_x  - \kk\Sk_{\kk}(x) \Sk_{\kk}(y) p_y}$ &  $ \displaystyle{ \Ck_{\kk}(r)\cos\phi\, p_r-\frac{\sin\phi}{\Sk_{\kk}(r)}\,p_\phi}$    \\[0.3cm]
$\pp_2$&$\displaystyle{ \sqrt{1+\kk\>q^2}\, p_2  }  $ & $\displaystyle {  \Ck_{\kk}(y) p_y}$  & $ \displaystyle{ \Ck_{\kk}(r)\sin\phi\, p_r+\frac{\cos\phi}{\Sk_{\kk}(r)}\,p_\phi}$  \\[0.4cm]
\hline
\\[-0.2cm]
 $ {\cal T}_\kk$&$ \displaystyle { \frac 12  (1+\kk \>q^2 )  (\>p^2+\kk(\>q\cdot\>p)^2  ) } \quad$ &  $\displaystyle{ \frac 12\left(\frac{p_x^2}{ \Ck_{\kk}^2(y)} +p_y^2  \right) }   $  &$ \displaystyle{ \frac 12\left(p_r^2+  \frac{p_\phi^2}{ \Sk_{\kk}^2(r)}   \right)  }   $ \\[0.4cm]
\hline
\\[-0.2cm]
 $J_{01}$&$ \displaystyle{  p_1+\kk (\>q\cdot\>p) q_1 } $ &  $  \displaystyle{  p_x }$  & $  \displaystyle{  \cos\phi\,p_r-\frac{\sin\phi}{ \Tk_\kk(r) }\, p_\phi}$\\[0.3cm]
$ J_{02}$&$  \displaystyle{ p_2+\kk (\>q\cdot\>p) q_2  }$&  $  \displaystyle{ \Ck_{\kk}(x)p_y+\kk \Sk_{\kk}(x)\Tk_{\kk}(y) p_x  }$  &$  \displaystyle{ \sin\phi\,p_r+\frac{\cos\phi}{\Tk_\kk(r)}\, p_\phi}$  \\[0.3cm]
$ J_{12}$&$\displaystyle{ q_1 p_2 - q_2 p_1 }  $ &    $  \displaystyle{ \Sk_{\kk}(x)p_y- \Ck_{\kk}(x)\Tk_{\kk}(y) p_x  }$  &  $  \displaystyle{  p_\phi}$\\[0.2cm]
\hline
\end{tabular}
\end{center}
 \end{table}

\end{landscape}

%%%%%%%%%%%%%%%%%%%%%%%%%%%%%%%%%%%%%%%%%%%%%%

%%%%%%%%%%%%%%%%%%%%%%%%%%%%%%%%%%%%%%%%%%%%%%

\section{Anisotropic oscillators on the Euclidean plane}
\label{sec4}

To start with, let us consider the   Hamiltonian determining the    anisotropic oscillator   with unit mass and   frequencies $ \omega_x$ and $ \omega_y$ on the 
 Euclidean plane  in Cartesian coordinates $(x,y)\in \mathbb R^2$ and    conjugate momenta  $(p_x,p_y)$:
\be
{H} = \frac12(p_x^2 + p_y^2) + \frac{ 1}{2}( \omega_x^2  x^2 + \omega_y^2y^2) .
\label{fa}
\ee
Clearly, this Hamiltonian   is always integrable due to its   separability in Cartesian coordinates so that  it Poisson-commutes with the (quadratic in the momenta) integrals of motion 
$$
{I}_x=\frac12 p_x^2  + \frac{ 1}{2} \omega_x^2  x^2,\qquad     {I}_y=\frac12 p_y^2  + \frac{ 1}{2}  \omega_y^2  y^2,
$$
which are not independent since
$$
{H} ={I}_x+{I}_y.
$$
 Furthermore,  it is also well-known that for  commensurate frequencies $\omega_x: \omega_y$ the Hamiltonian (\ref{fa}) provides a superintegrable system~\cite{Jauch,Stefan,Tempesta}, in such a manner that    an  ``additional"  (in general higher-order in the momenta) integral of motion does exist. 
 
The (super)integrability properties of the commensurate oscillator will be sketched by following the approach given in~\cite{Kuruannals,KuruProcs},  which is  based on a classical factorization  formalism  (see~\cite{david,Kuru1,kuru12,Kuru3,lissajous} and references therein).  If we denote
\be
 \omega_x=\gamma \omega_y ,\qquad  \omega_y=\omega ,\qquad  \gamma\in  \mathbb R^+/\{0\},
 \label{fb}
\ee
then  $H$ (\ref{fa}) can be written in terms  of the   parameter $\gamma$ and   frequency $\omega$ as
\begin{equation}
H = \frac12(p_x^2 + p_y^2) + \frac{\omega^2}{2} \left( (\gamma x)^2 + y^2 \right) .
\label{fc}
\end{equation}
Next  we introduce     new canonical variables
\be
\xi=\gamma x,\qquad  p_\xi =p_x/\gamma,\qquad \xi\in  \mathbb R,
\label{fd}
\ee
giving rise to
\begin{equation}
H= \frac12\, p_y^2 + \frac{\omega^2}{2} \,  y^2 
 +  \gamma^2\left(\frac12\, p_\xi^2 + \frac{\omega^2}{2\gamma^2} \,  \xi^2\right) .
 \label{fe}
\end{equation}
Therefore we obtain two 1D Hamiltonians $H^\xi$ and $H^y$  given by
\be
H^\xi = \frac12\, p_\xi^2 + \frac{\omega^2}{2\gamma^2} \xi^2  ,\qquad  H^y = \frac12\, p_y^2 + \frac{\omega^2}{2} y^2  ,\qquad H= H^y 
 +  \gamma^2 H^\xi , 
\label{ff}
\ee
which are two integrals of the motion for $H$. The 1D Hamiltonian $H^\xi $ (\ref{ff}) can then be factorized in terms of    ``ladder    functions" $B^\pm$ as
\begin{equation}
H^\xi =B^+ B^-,\qquad B^{\pm} ={\mp}\frac{i}{\sqrt{2}}\, p_{\xi}+
\frac{1}{\sqrt2} \frac{\omega}{\gamma}\,  \xi \, ,
\label{fg}
\end{equation}
fulfilling
\be
\{H^\xi ,B^{\pm} \}=\mp i\, \frac{\omega}{\gamma} \,B^{\pm} , \qquad  \{ B^-,B^+ \}= - i \,\frac{\omega}{\gamma} .
 \nonumber
\ee
The remaining  1D Hamiltonian $H^y $ (\ref{ff}) can also be factorized through    ``shift functions'' $A^\pm$   in the form
\begin{equation}
H^y= A^+ A^- ,\qquad A^{\pm}=\mp \frac{i}{\sqrt{2}}\,p_{y}-\frac{\omega}{\sqrt{2}}\,  y \, ,
\label{fh}
\end{equation}
so that
 \be
\{H^y ,A^{\pm} \}=\pm  i  {\omega}  A^{\pm} , \qquad  \{ A^-,A^+ \}=   i {\omega}  .
\nonumber
\ee
Notice that the sets of   functions $(H^\xi,B^\pm,1)$ and  $(H^y,A^\pm,1)$ span        a  Poisson--Lie   algebra isomorphic to the  harmonic oscillator Lie algebra $\mathfrak{h}_4$.  Hence, the 2D  Hamiltonian (\ref{fe}) can finally be  expressed in terms of the above ladder and shift functions as
\begin{equation}
H= A^+ A^- +\gamma^2 B^+ B^-,\qquad \{H ,B^{\pm}\}=\mp i {\gamma} {\omega} B^{\pm}   ,\qquad \{H ,A^{\pm}\}=\pm{ i \omega} A^{\pm}\,.
\label{fj}\nonumber
\end{equation}

The remarkable fact  now is that if we consider a rational value for $\gamma$,
\be
\gamma  = \frac{\omega_x}{\omega_y}=\frac{m}{n}   , \qquad m,n\in \mathbb N^\ast ,
\label{fjh}
\ee
we obtain  two additional complex
constants of the motion $X^{\pm}$ for $H$  (\ref{fe})  
\begin{equation}
X^{\pm}=(B^{\pm})^n (A^{\pm})^{m}\, , \qquad \bar{X}^+= {X}^- ,  
\label{fk}
\end{equation}
which are of $(m+n)$th-order in the momenta. Real-valued integrals of the motion can be defined through the expressions
\be
X= \frac 12(X^+ + X^-),\qquad Y= \frac 1{2i} (X^+ - X^-).
\label{fl}
\ee
The final result can be summarized as follows~\cite{Kuruannals,KuruProcs}.
 
\begin{enumerate}
\item The Hamiltonian $H$ (\ref{fe}) is  always integrable for any value of the real parameter $\gamma$, since it is endowed with a quadratic constant of the motion given by either $H^\xi$ or $H^y$ (\ref{ff}).\\
\item When $\gamma=m/n$ is a rational parameter (\ref{fjh}), the Hamiltonian (\ref{fe}) defines a  superintegrable  anisotropic oscillator with commensurate frequencies $\omega_x:\omega_y$ and the additional constant of the motion is given by either $X$ or $Y$ in (\ref{fl}). The sets $(H,H^\xi,X)$ and $(H,H^\xi,Y)$ are formed by three functionally independent functions.\\
\item When $(m+n)$ is even, the highest order constant  of the motion in the momenta is  $X$ being of $(m+n)$-degree,  while  $Y$ is
 of lower order $(m+n-1)$.  When  $(m+n)$ is odd, the highest $(m+n)$-degree integral is $Y$ while  $X$ is
 of lower order $(m+n-1)$.
 \end{enumerate}

 It is worth recalling that the Hamiltonian (\ref{fa}) can be enlarged by adding two Rosochatius (or Smorodinski-Winternitz)  terms as
\be
{H}_\lambda = \frac12(p_x^2 + p_y^2) + \frac{ 1}{2}( \omega_x^2  x^2 + \omega_y^2y^2) +\frac{\lambda_1}{x^2}+\frac{\lambda_2}{y^2},
\label{faa}
\ee
where $\lambda_1$ and $\lambda_2$ are real parameters, which provide centrifugal barriers when both constants are positive ones. In the 3D case, the resulting Hamiltonian, called ``caged anisotropic oscillator", has been solved in~\cite{VerrierOscillator} (for both classical and quantum systems),  and the  general $N$D case has been fully studied in~\cite{Tempesta}. Despite the introduction of the $\lambda_i$-potentials, the  Hamiltonian  (\ref{faa}) is again (maximally) superintegrable for commensurate frequencies (in any dimension).

We  also remark that any    $m:n$  oscillator  (labelled by $\gamma$) is equivalent to the  $n:m$  one (with $1/\gamma$)   via the interchanges $x\leftrightarrow y$  and $\gamma\omega \leftrightarrow \omega$.
Consequently, according to the above statement the only anisotropic oscillators which are quadratically superintegrable correspond to the cases with $\gamma=1$ and $\gamma=2$ (so also $\gamma=1/2$), in agreement with the classifications on superintegrable Euclidean systems given in~\cite{RS,Ev90Pra,KaWiMiPo99}. In the sequel, we will illustrate the previous general results by working out these two particular cases.

%%%%%%%%%%%%%%%%%%%%%%%%

\subsection{The $\gamma=1$ or 1:1   (isotropic)  oscillator}
\label{sec41}

 We   set $m=n=1$ so that the relations (\ref{fb}) and (\ref{fd}) simply give $\omega_x=\omega_y=\omega$ and  $\xi=x$ and $p_\xi=p_x $. Thus we recover the isotropic oscillator 
  \be
{H}^{1:1} = \frac12(p_x^2 + p_y^2) + \frac{ \omega^2 }{2}( x^2 + y^2) ,
  \label{ffmm}
  \ee
  and the constants of the motion (\ref{fl}) reduce to
\be
X=-\tfrac 12 (p_x p_y +\omega^2 x y) ,\qquad Y =-\tfrac 12 \omega (x p_y - y p_x)   .
\label{fm}
\ee
Since $(m+n)=2$ is even, we find a 
  quadratic integral $X$, which  is one of the components of the Demkov--Fradkin tensor~\cite{Demkov,Fradkin}, and  a  first-order one $Y$ which is proportional to the angular momentum 
\be
\mathcal{J} = x p_y - y p_x .
\label{fn}
\ee

  We recall that if we add the two ``centrifugal"  $\lambda_i$-terms (\ref{faa}), then  we get  the 2D version  of the so-called Smorodinsky--Winternitz system~\cite{FrMaSmUW65} which has been widely studied  (see, e.g.,~\cite{BaHeSantS03,LetterBH,KaWiMiPo99,Evans90,Evans91,GrPoSi95a} and references therein).

%%%%%%%%%%%%%%%%%%%%%%%%

\subsection{The $\gamma=2$ or 2:1   oscillator}
\label{sec42}

If we take $m=2$ and $n=1$, then $\omega_x=2\omega_y=2\omega$, $\xi=2 x$ and $p_\xi =p_x/2$. The    Hamiltonian (\ref{fe})  and the  integrals (\ref{fl}) turn out to be 
\bea
&& H^{2:1} = \frac12\, p_y^2 + \frac{\omega^2}{2} \, y^2 
 +  4\left(\frac12\, p_\xi^2 + \frac{\omega^2}{8}\, \xi^2\right) =\frac12(p_x^2 + p_y^2) + \frac{\omega^2}{2} \left( 4 x^2 + y^2 \right) ,\nonumber\\[2pt]
 && X=-\frac{\omega}{4\sqrt{2}} \left( p_y ( \xi p_y - 4 y p_\xi) - \omega^2 \xi y^2  \right)= -\frac{\omega}{2\sqrt{2}} \left( p_y \mathcal{J}  - \omega^2 x y^2  \right) , \label{fp} \\[2pt]
 &&Y=  \frac{1}{2\sqrt{2}} \left( p_\xi p_y^2 +   \omega^2 y ( \xi p_y -   y p_\xi)    \right)=  \frac{1}{4\sqrt{2}} \left( p_x p_y^2 +   \omega^2 y ( 4 x p_y -   y p_x)    \right) .
\nonumber
\eea
In this case $(m+n)=3$ is odd, so that we find a cubic integral $Y$ and a   quadratic one $X$, which  involves the angular momentum $\mathcal{J}$ (\ref{fn}); the latter is the integral of the motion which is usually considered in the literature (see e.g.~\cite{RS,WolfBoyer}) and shows that 
  the 2:1 oscillator can be regarded as a superintegrable system with {\em quadratic} constants of the motion. 

Notice that if we add a single Rosochatius-Winternitz potential  $\lambda_2/y^2$ (by setting $\lambda_1=0$) the generalized system remains quadratically superintegrable~\cite{RS}, but if both $\lambda_i$-terms are introduced, the additional integral turns out to be of sixth-order in the momenta~\cite{Tempesta}.

%%%%%%%%%%%%%%%%%%%%%%%%%%%%%%%%%%%%%%%%%%%%%%%

\section{Anisotropic oscillators on ${\>S}^{2}$ and ${\>H}^{2}$}
\label{sec5}

Let us recall that the classification of all possible superintegrable systems on ${\mathbf S}^2$ and ${\mathbf H}^2$ with quadratic integrals of the motion was performed in~\cite{RS}, where two curved superintegrable oscillator potentials were found:
\begin{itemize}

\item The isotropic Higgs oscillator~\cite{Higgs}, whose Euclidean limit is the 1:1 isotropic oscillator (\ref{ffmm}).\\

\item
The curved version of the superintegrable Euclidean 2:1 oscillator (\ref{fp}). 

\end{itemize}

The aim of this Section is to present the construction of the constant-curvature Hamiltonian analogue $H_\kk$ of the anisotropic Euclidean Hamiltonian $H$ (\ref{fc}) with arbitrary commensurate frequencies. The idea is to introduce appropriately the curvature parameter $\kk$  by requiring to keep the same (super)integrability properties as given in the previous Section for (\ref{fc}) (or (\ref{fe})) and, simultaneously, allowing  for a smooth and well-defined flat limit $\kk\to 0$ of the curved Hamiltonian and its constants of the motion. 

This result has been achieved in~\cite{Kuruannals} by using the geodesic parallel variables described in Section~\ref{sec22},  so with     kinetic term  ${\cal T}_\kk$  given by (\ref{ed}). Explicitly, the curved Hamiltonian ${H}_{\kappa} $ has been shown to be of the form
\begin{equation}
{H}_{\kappa} =  {\cal T}_\kk + {U}^\gamma_\kk = \frac{1}{2} \left(\frac{p_{x}^2}{ \Ck_\kk^2(y)}+p_{y}^2\right)+\frac{\omega^2}{2} 
\left(\frac{ \Tk_{\kappa}^2(\gamma x)}{ \Ck_{\kappa}^2(y)}+\Tk_{\kappa}^2(y)\right) .
\label{ga}
\end{equation}
By taking into account  (\ref{eedd}) and the presence of the $\kk$-tangent
$\Tk_{\kappa}(\gamma x)$ in the potential $ {U}^\gamma_\kk$, we find that  the domain of the geodesic parallel coordinates $(x,y)$ is restricted to
\bea
 {\mathbf S}^2\ (\kk>0) :&&     -\frac{\pi}{2\sqrt{\kk}}< \gamma x< \frac{\pi}{2\sqrt{\kk}} ,\quad   -\frac{\pi}{2\sqrt{\kk}}< y< \frac{\pi}{2\sqrt{\kk}} ,  \quad \gamma\ge \frac 12  . \nonumber \\
 {\mathbf H}^2\ (\kk<0) :&&  x,y \in \mathbb{R} ,\quad \gamma\in  \mathbb R^+/\{0\}.
\nonumber
\eea

Now we assume that  $\kk\ne  0$ and since $ 1+\kk  \Tk_{\kappa}^2(u)=1/\Ck_{\kappa}^2(u)$ we rewrite  the  Hamiltonian ${H}_{\kappa}$ (\ref{ga})  as 
\begin{equation}\label{gb}
H_{\kappa} = 
\frac{p_{y}^2}{2}+\frac{1}{\Ck_{\kappa}^2(y)}\left(\frac{p_{x}^2}{2}+
\frac{\omega^2}{2 \kappa \Ck_{\kappa}^2(\gamma x)}\right)-
\frac{\omega^2}{2\kappa} ,\qquad \kk\ne 0.
\end{equation}
Next we introduce the canonical variables $(\xi,p_\xi)$  (\ref{fd}) finding that
\begin{equation}
H_{\kappa} = \frac{p_{y}^2}{2}+\frac{\gamma^2}{\Ck_{\kappa}^2(y)}\left(\frac{p_{\xi}^2}{2}+\frac{\omega^2}{2 \kappa  \gamma^2 \Ck_{\kappa}^2(\xi)}\right)-
\frac{\omega^2}{2  \kappa}\, ,\qquad \kk\ne 0.
\nonumber
\end{equation}
And the curved Hamiltonian (\ref{gb}) is finally expressed as
\begin{equation}\label{gc}
H_{\kappa}  = \frac{p_{y}^2}{2}+\frac{\gamma^2 H_{\kappa}^\xi}{\Ck_{\kappa}^2(y)}-
\frac{\omega^2}{2\kappa} ,\qquad H_{\kappa}^\xi = \frac{p_{\xi}^2}{2}
+\frac{\omega^2}{2\kappa\gamma^2\Ck_{\kappa}^2(\xi)} ,  \qquad \kk\ne 0,
\end{equation}
where $H_{\kappa}^\xi $ is  a constant of the   motion. Notice that  in this form, the 1D Hamiltonians $H_{\kappa}^\xi $  and  $H_{\kappa} $ 
 correspond to   P\"oschl--Teller systems~\cite{Kuru1}. Consequently,   $H_{\kappa}$ determines an integrable system for any value of $\omega$ and $\gamma$.  

In the sequel,
we factorize  the 1D Hamiltonians $H_{\kappa}^\xi$ and  $H_{\kappa}$ (\ref{gc}) (see~\cite{KuruProcs,Kuruannals,Kuru1} for details).   By one hand, the Hamiltonian $H_{\kappa}^\xi$ is factorized in terms of   ladder  functions as 
   \be\label{gd}
H_{\kappa}^\xi=B_\kk^+B_\kk^- + \frac{\omega^2}{2\kappa\gamma^2},\qquad B_\kk^{\pm} ={\mp}\frac{i}{\sqrt{2}}\Ck_{\kappa}(\xi)\, p_{\xi}+
\frac{\mm}{\sqrt{2}} \Sk_{\kappa}(\xi),
\ee
where $\mm$ is a constant of the motion  defined by
\be
\mm(p_{\xi},\xi) := \sqrt{2\kappa H_{\kappa}^\xi} \, .
\label{ge}
\ee
Thus we get    the  Poisson algebra 
\begin{equation} 
\{H_{\kappa}^\xi,B_\kk^{\pm}\}= \mp i\,\mm\,B_\kk^{\pm} ,\qquad \{ B_\kk^-,B_\kk^+\}=- i  \,\mm \, ,
\nonumber
\end{equation}
so that
\begin{equation} 
\{H_{\kappa} ,B_\kk^{\pm}\}=\mp  i\,\frac{\gamma^2 \mm}{\Ck_{\kappa}^2(y)}\,B_\kk^{\pm} .
\nonumber
\end{equation}
On the other hand,    $H_{\kappa}$ is factorized by means  of   shift  functions in the form
 \begin{equation}\label{gh}
H_{\kappa} =A_\kk^{+} A_\kk^{-}+\frac{1}{2\kappa}  \left(\gamma^2\mm^2-\omega^2\right) ,\qquad A_\kk^{\pm}=\mp \frac{i}{\sqrt{2}}\,p_{y}-\frac{\gamma \mm}{\sqrt{2} }\Tk_{\kappa}(y) ,
\end{equation}
 closing on  the Poisson algebra
\begin{equation} 
\{H _{\kappa},A_\kk^{\pm}\}=\pm i\,\frac{\gamma\mm}{\Ck_{\kappa}^2(y)}\,A_\kk^{\pm} ,\qquad \{ A_\kk^-,A_\kk^+\}=i\,\frac{\gamma\mm}{\Ck^2_{\kappa}(y)} .
\nonumber
\end{equation}

As in the Euclidean case, when $\gamma$ takes a rational value (\ref{fjh}), two additional complex integrals of motion arise for $H_\kk$ (\ref{ga}) (under the change of variable  (\ref{fd})), namely
\be
 X_\kk^{\pm}=(B_\kk^{\pm})^n (A_\kk^{\pm})^{m} ,\qquad   \bar{X}_\kk^+= {X}_\kk^- .
\label{gi}
\ee
Nevertheless,   in order to obtain real constants of the motion,  $\xx_\kk$ and $\yy_\kk$, we are now led to distinguish between two situations~\cite{lissajous} (due to the presence of powers of $\mm$ (\ref{ge}) in (\ref{gi})):
\be
\begin{aligned}
&\mbox{When $m+n$ is even:}\quad  X_\kk^\pm = \pm i \, \mm \yy_\kk  + \xx_\kk . \\
&\mbox{When  $m+n$ is odd:} \quad X_\kk^\pm = \mm\xx_\kk  \pm i \,\yy_\kk . 
\end{aligned}
\label{gj}
\ee

 Summing up, the generalization to $\mathbf S^2$ and $\mathbf H^2$ of the anisotropic oscillator can be stated as follows~\cite{Kuruannals}.

\begin{enumerate}
\item  For any value of $\gamma$,   the Hamiltonian $H_\kk$ (\ref{ga}) always determines an   integrable anisotropic curved oscillator on ${\mathbf S}^2$   and $ {\mathbf H}^2$, with  quadratic constant of motion   given by   $H^\xi_\kk$   (\ref{gc}).\\
\item When $\gamma$ is a rational parameter (\ref{fjh}),   the Hamiltonian $H_\kk$    defines   a  superintegrable  anisotropic curved oscillator  and the additional constant of  the motion  is given by either $\xx_\kk$ or $\yy_\kk$ in (\ref{gj}). The   sets $(H_\kk,H_\kk^\xi,\xx_\kk)$ and $(H_\kk,H_\kk^\xi,\yy_\kk)$ are formed by three functionally independent functions.\\
\item The integrals $\xx_\kk$ and $\yy_\kk$ are polynomial in the momenta, whose degrees are $(m+n)$ and $(m+n-1)$ when $(m+n)$ is even, and $(m+n-1)$ and $(m+n)$ when $(m+n)$ is odd, respectively.
\end{enumerate}

Some remarks are in order. Firstly,   although the  (flat) Euclidean  limit $\kk\to 0$ is precluded for $H_\kk$   written in the forms (\ref{gb}) and  (\ref{gc}), it is actually well-defined in all the remaining expressions. To perform the contractions one has to take into account   the following flat limit  of the integrals $H_{\kappa}^\xi$ (\ref{gc}) and $\mm$ (\ref{ge})   
\be
\lim_{\kk\to 0} \kk H_{\kappa}^\xi=\frac{\omega^2}{2\gamma^2},\qquad \lim_{\kk\to 0} \mm=\frac{\omega}{\gamma} .
\label{gk}
\ee
Therefore, it can be easily checked that  when $\kk\to 0$,  the curved  Hamiltonian $H_{\kappa}$ (\ref{ga})  reduce to $H$  (\ref{fc}), the curved ladder functions $B_{\kappa}^\pm$    (\ref{gd})   to $B^\pm$    (\ref{fg}), the curved shift functions $A_{\kappa}^\pm$    (\ref{gh})   to $A^\pm$    (\ref{fh}),   and so the curved integrals $X_\kk^{\pm}$ (\ref{gi}) to $X^{\pm}$ (\ref{fk}).

Secondly, as in the Euclidean system (\ref{faa}),  the curved  Hamiltonian $H_{\kappa}$ (\ref{ga})  can be generalized by    adding two curved Rosochatius--Winternitz  potentials  which in ambient coordinates (\ref{eb})  adopt a very simple expression~\cite{BaHeSantS03,LetterBH}. Explicitly, the corresponding potential reads
\be
{U}^\gamma_{\kk,\lambda} = {U}^\gamma_\kk+ \frac{\lambda_1}{x_1^2}+ \frac{\lambda_2}{x_2^2}=
\left(\frac{ \Tk_{\kappa}^2(\gamma x)}{ \Ck_{\kappa}^2(y)}+\Tk_{\kappa}^2(y)\right)+ \frac{\lambda_1}{\Sk^2_{\kk }(x)\Ck^2_{\kk }(y)}+ \frac{\lambda_2}{\Sk^2_{\kk }(y)} .
\label{gl}
\ee
Then the corresponding Hamiltonian ${H}_{\kappa,\lambda} =  {\cal T}_\kk + {U}^\gamma_{\kk,\lambda}$ can be written as (with $\kk\ne 0$)
\begin{equation} 
{H}_{\kappa,\lambda} =
\frac{p_{y}^2}{2}+ \frac{\lambda_2}{\Sk^2_{\kk }(y)}+\frac{1}{\Ck_{\kappa}^2(y)}\left(\frac{p_{x}^2}{2}+
\frac{\omega^2}{2 \kappa \Ck_{\kappa}^2(\gamma x)}  + \frac{\lambda_1}{\Sk^2_{\kk }(x) }\right)-
\frac{\omega^2}{2\kappa} , 
\nonumber
\end{equation}
to be compared with (\ref{gb}). Consequently, ${H}_{\kappa,\lambda}$ defines an integrable system for any value of $\omega$, $\gamma$, $\lambda_1$ and $\lambda_2$.

Now it could be expected that if $\gamma$ is a rational number, ${H}_{\kappa,\lambda}$ should be again superintegrable but, to the best of our knowledge, this property has not been proven in general  (except for $\gamma=1$). We also point out that each $\lambda_i$-term gives rise to a centrifugal barrier on  $\mathbf H^2$  when $\lambda_i>0$, as in the Euclidean system but, surprisingly enough, both $\lambda_1$- and $\lambda_2$-potentials can be interpreted as noncentral 1D curved   oscillators on $\mathbf S^2$ with centres at the points $O_1=(0,1,0)$ and $O_2=(0,0,1)$, respectively~\cite{ran,ran1,BaHeSantS03,CRMVulpi,BaHeMu13,ran2} (see   Figure~\ref{figure2}).

Thirdly,   it can be seen  from $H_{\kappa}$ (\ref{ga}) that, in general,    $U_\kk^{\gamma} $ and $U_\kk^{1/\gamma} $   determine two different  systems,
 in contradistinction with  the Euclidean case (recall that the equivalence was provided by the interchanges $x\leftrightarrow y$  and $\gamma\omega \leftrightarrow \omega$). However   when $\kk=0$ both potentials reduce to equivalent Euclidean potentials. This   clearly illustrates the fact that given a flat Hamiltonian system there could be not a single but several curved generalizations (or curvature integrable deformations) which would be non-equivalent in the sense that no canonical change of variables  exist between them.
 
Fourthly,   according to the results previously presented, the only anisotropic curved oscillators which are quadratically superintegrable correspond to the same values of $\gamma$ as in the Euclidean system~\cite{RS,Kalnins1,Kalnins2}: $\gamma=1$,  $\gamma=2$ and (now the non-equivalent)  $\gamma=1/2$.  
In what follows we shall present the corresponding results for the these three cases.

  And finally, we recall that  other integrable anisotropic oscillators on  the spheres and hyperbolic spaces can be found in~\cite{kalnins,Saksida,Nersessian,Marquette1} (see also references therein).

%%%%%%%%%%%%%%%%%%%%%%%%%%%%%%%%%%%%%%%%%%%%%%%

\subsection{The $\gamma=1$ or  1:1 curved (isotropic) oscillator}
\label{sec51}

This case  is also known as the Higgs oscillator~\cite{Higgs,Leemon} and it has been widely studied in the literature (see~\cite{RS,Santander6,BaHeMu13,Pogoa,Nersessian1,Ranran,Annals09} and references therein).

We set   $\gamma=m=n=1$ so that  $\xi=x$ and $p_\xi=p_x $. The Hamiltonian  $H_\kk$ (\ref{ga})  reduces to
$$
{H}_{\kappa}^{1:1} =   \frac{1}{2} \left(\frac{p_{x}^2}{ \Ck_\kk^2(y)}+p_{y}^2\right)+\frac{\omega^2}{2} 
\left(\frac{ \Tk_{\kappa}^2(  x)}{ \Ck_{\kappa}^2(y)}+\Tk_{\kappa}^2(y)\right)  = \frac{p_{y}^2}{2}+\frac{  H_{\kappa}^x}{\Ck_{\kappa}^2(y)}-
\frac{\omega^2}{2\kappa}  ,
$$
where the quadratic  integral $H_{\kappa}^x\equiv H_{\kappa}^\xi$ (\ref{gc}) is given by 
\be
H_{\kappa}^x= \frac{p_{x}^2}{2}
+\frac{\omega^2 }{2\kappa\Ck_{\kappa}^2(x)} .
\nonumber
\ee

Since $(m+n)=2$ is even  the constants of the motion (\ref{gj}) read
\bea
&& \xx_\kk=-\frac 12\left(  \Ck_{\kappa} (x) p_xp_y+\mm^2 \Sk_{\kappa} (x)\Tk_{\kappa} (y)\right) ,\nonumber\\[2pt]
&&\yy_\kk=-\frac 12\,\left(  \Sk_{\kappa} (x) p_y- \Ck_{\kappa} (x)\Tk_{\kappa} (y) p_x\right) .
\nonumber
\eea
The integral $\yy_\kk$ is proportional to the (curved) angular momentum $\ang_\kk$ which in geodesic parallel and polar variables is given by~\cite{RS,conf}
\be
\ang_\kk =\Sk_{\kappa} (x) p_y- \Ck_{\kappa} (x)\Tk_{\kappa} (y) p_x= p_\phi .
\nonumber
\ee
The flat limit $\kk\to 0$ of all the above expressions leads to  the results of the Euclidean isotropic oscillator given in Section~\ref{sec41}. In particular,
provided that    $\mm \to \omega$ (\ref{gk}), the integrals $\xx_\kk$, 
  $\mm \yy_\kk$   and the curved angular momentum $\ang_\kk$ reduce to   (\ref{fm})  and (\ref{fn}).

Notice that the potential of ${H}_{\kappa}^{1:1}$ is expressed in terms of ambient and geodesic polar coordinates (\ref{eb}) as
$$
{U}_{\kappa}^{1:1} =  \frac{\omega^2}2\left( \frac{x_1^2+x_2^2}{x_0^2} \right) = \frac{\omega^2}2  \Tk^2_\kk(r) .
$$

If we consider  the two $\lambda_i$-potentials (\ref{gl}), we recover the    curved Smorodinsky--Winternitz system
$$
{U}_{\kappa,\lambda}^{1:1} =  \frac{\omega^2}2\left( \frac{x_1^2+x_2^2}{x_0^2} \right) + \frac{\lambda_1}{x_1^2}+ \frac{\lambda_2}{x_2^2},
$$
which is known to be quadratically superintegrable~\cite{RS,BaHeSantS03,CRMVulpi,LetterBH,BaHeMu13,Pogosyan1}.

%%%%%%%%%%%%%%%%%%%%%%%%%%%%%%%%%%%%%%%%%%%%%%%

\subsection{The $\gamma=2$ or  2:1 curved oscillator}
\label{sec52}

We set $\gamma=m=2$ and $n=1$ so that $\xi=2x$ and $p_\xi =p_x/2 $.   The Hamiltonian $H_\kk$ (\ref{ga})  reads
\be
{H}_{\kappa}^{2:1}  =   \frac{1}{2} \left(\frac{p_{x}^2}{ \Ck_\kk^2(y)}+p_{y}^2\right)+\frac{\omega^2}{2} 
\left(\frac{ \Tk_{\kappa}^2( 2 x)}{ \Ck_{\kappa}^2(y)}+\Tk_{\kappa}^2(y)\right)  = \frac{p_{y}^2}{2}+\frac{ 4 H_{\kappa}^\xi}{\Ck_{\kappa}^2(y)}-
\frac{\omega^2}{2\kappa}  ,
\label{pb}\nonumber
\ee
where 
\be
H_{\kappa}^\xi= \frac{p_{\xi}^2}{2}
+\frac{\omega^2 }{8\kappa\Ck_{\kappa}^2(\xi)}= \frac{p_{x}^2}{8}
+\frac{\omega^2 }{8\kappa\Ck_{\kappa}^2(2x)} .
\nonumber
\ee

Now $(m+n)=3$ is odd so that the constants of the motion (\ref{gj}) turn out to be
\bea
&& \xx_\kk=
-\frac {1}{2\sqrt{2}}\left( \left[   \Sk_{\kappa} (2x)p_y-2  \Ck_{\kappa} (2x)  \Tk_{\kappa} (y) p_x  \right] p_y     - 4 \mm^2   \Sk_{\kappa} (2x)   \Tk^2_{\kappa} (y)\right)  ,\nonumber
 \\[2pt]
&&\yy_\kk=
\frac {1}{4\sqrt{2}}\left(      
     \!   \Ck_{\kappa} (2x)    p_x p_y^2 + 4 \mm^2\Tk_{\kappa} (y) \left[ 2\Sk_{\kappa} (2 x)p_y -  \Ck_{\kappa} (2 x)     \Tk_{\kappa} (y)  p_x   \right]          \right),
\nonumber
\eea
that is $ \xx_\kk$  is quadratic in the momenta, while  $ \yy_\kk$  is cubic; this means that ${H}_{\kappa}^{2:1} $ is a quadratically superintegrable system.
The limit $\kk\to 0$  (\ref{gk}) gives $\mm \to \omega/2$, hence the Hamiltonian ${H}_{\kappa}^{2:1} $ and the integrals $\mm \xx_\kk$ and $\yy_\kk$ reduce to (\ref{fp}), thus reproducing the results of Section~\ref{sec42}.

 In terms of ambient and geodesic polar coordinates (\ref{eb}), the potential of ${H}_{\kappa}^{2:1} $ adopts the (cumbersome) expressions
\bea
&&\!\!\!{U}_{\kappa}^{2:1} =\frac{\omega^2}{2} 
\left(   \frac{4 x_0^2x_1^2}{(x_0^2+\kk x_1^2)(x_0^2-\kk x_1^2)^2}+ \frac{x_2^2}{(1-\kk x_2^2)} \right)    \nonumber\\[2pt]
&&\qquad = \frac{\omega^2}{2} \left( \frac{4\Tk^2_\kk(r)\cos^2\phi}{\left(1-\kk \Sk^2_\kk(r)\sin^2\phi\right)\left(1-\kk \Tk^2_\kk(r)\cos^2\phi\right)^2} +  \frac{\Sk^2_\kk(r)\sin^2\phi}{ 1-\kk \Sk^2_\kk(r)\sin^2\phi }     \right) 
\nonumber
\eea
the latter is the one formerly introduced in~\cite{RS}.

In this case  it is only possible to add a single    $\lambda_i$-potential  (\ref{gl}) keeping the quadratic superintegrability of the system~\cite{RS}
$$
{U}_{\kappa,\lambda}^{2:1} ={U}_{\kappa}^{2:1} +\frac{\lambda_2}{x_2^2} ,
$$ 
which has been studied in detail in~\cite{BaHeMu13,BaBlHeMu14}. In this respect, we remark that an equivalent superintegrable system can be obtained by interchanging the ambient coordinates $x_1\leftrightarrow x_2$ (so the   role of the geodesics $l_1\leftrightarrow l_2$ in Section~\ref{sec22}) which means that the   geodesic parallel coordinates are mapped as $ (x,y)\to  (y',x')$, that is,
$$
 x_0 =\Ck_{\kk}(x')\Ck_{\kk}(y') ,\qquad  x_1 =\Sk_{\kk }(x')  ,\qquad    x_2=\Ck_{\kk }(x')\Sk_{\kk }(y')  .
 $$
 In fact, the coordinates $(x',y')$ are just the so-called geodesic parallel coordinates of type II~\cite{conf}.
These transformations provide  the equivalent system
\bea
&& {H'}_{\kappa}^{2:1} ={\cal T}_\kk +\frac{\omega^2}{2} 
\left(   \frac{x_1^2}{(1-\kk x_1^2)}+  \frac{4 x_0^2x_2^2}{(x_0^2+\kk x_2^2)(x_0^2-\kk x_2^2)^2}\right)    \nonumber\\[2pt]
&&\qquad \quad = \frac{1}{2} \left( p_{x'}^2  +  \frac{p_{y'}^2}{ \Ck_\kk^2(x')}\right)+ \frac{\omega^2}{2} 
\left(\Tk_{\kappa}^2(x')+ \frac{\Tk_{\kappa}^2( 2y')}{ \Ck_{\kappa}^2(x')}\right)  .
\label{xzy}
\eea
This is exactly the expression for the 2:1 curved oscillator   considered in~\cite{BaHeMu13,BaBlHeMu14}.

%%%%%%%%%%%%%%%%%%%%%%%%%%%%%%%%%%%%%%%%%%%%%%%

\subsection{The  $\frac 12$:1 curved oscillator}
\label{sec53}

We set  $\gamma=1/2$,  $m=1$ and $n=2$, so that $\xi= x/2$, $p_\xi =2 p_x $. Thus the   Hamiltonian $H_\kk$ (\ref{ga})  is 
\be
{H}_{\kappa}^{\frac 12 :1}  =   \frac{1}{2} \left(\frac{p_{x}^2}{ \Ck_\kk^2(y)}+p_{y}^2\right)+\frac{\omega^2}{2} 
\left(\frac{ \Tk_{\kappa}^2( \frac x2)}{ \Ck_{\kappa}^2(y)}+\Tk_{\kappa}^2(y)\right)  = \frac{p_{y}^2}{2}+\frac{  H_{\kappa}^\xi}{4\Ck_{\kappa}^2(y)}-
\frac{\omega^2}{2\kappa}\,  ,
\nonumber
\ee
where 
\be
H_{\kappa}^\xi= \frac{p_{\xi}^2}{2}
+\frac{2\omega^2 }{ \kappa\Ck_{\kappa}^2(\xi)}= 2  {p_{x}^2} 
+\frac{2\omega^2 }{ \kappa\Ck_{\kappa}^2(\frac x2)} .
\nonumber
\ee

The sum $(m+n)=3$ is again odd, and  the additional integrals (\ref{gj})   read
\bea
&& \xx_\kk=
-\frac {1}{4\sqrt{2}}\left( 4\left[   \Sk_{\kappa} (x)p_y-   \Ck^2_{\kappa} (\tfrac x2)  \Tk_{\kappa} (y) p_x  \right] p_x     + \mm^2   \Sk^2_{\kappa} (\tfrac x2)   \Tk_{\kappa} (y)\right)  ,\nonumber\\[2pt]
&&\yy_\kk=
\frac {1}{2\sqrt{2}}\left(      
      4  \Ck^2_{\kappa} (\tfrac x2)    p^2_x p_y - \mm^2  \left[  \Sk^2_{\kappa} (\tfrac x2)p_y -  \Sk_{\kappa} (  x)     \Tk_{\kappa} (y)  p_x   \right]          \right),
  \nonumber
\eea
which shows that ${H}_{\kappa}^{\frac 12 :1}$ is again a quadratically superintegrable system.

As a consequence, when both Hamiltonians ${H}_{\kappa}^{2 :1}$ and ${H}_{\kappa}^{\frac 12 :1} $ are considered altogether, one finds a particular issue where the curvature-deformation approach gives rise to two non-equivalent systems starting from the common  ``seed"  given by the Euclidean system ${H}^{2 :1}\simeq   {H}^{\frac 12 :1} $ (\ref{fp})   described in Section~\ref{sec42}.
Therefore, the plurality of possible integrable curved generalizations of a given Euclidean system becomes evident, and a deeper analysis of the curved system ${H}_{\kappa}^{\frac 12 :1} $ seems to be needed, since --to the best of our knowledge-- it has not been appropriately considered in the literature so far.

%%%%%%%%%%%%%%%%%%%%%%%%%%%%%%%%%%%%%%%%%%%%%%%

\section{Integrable H\'enon--Heiles systems}
\label{sec6}

By making use of the results described in the previous Sections, our aim now will be to present the generalization to the 2D sphere  and  the hyperbolic space of the integrable H\'enon--Heiles Hamiltonian  given by
\be
\mathcal{H}=\dfrac{1}{2}(p_{1}^{2}+p_{2}^{2})+ \Omega \left(  q_{1}^{2}+ 4 q_{2}^{2}\right) +\alpha \left(
q_{1}^{2}q_{2}+2 q_{2}^{3}\right) ,
\label{KdVflat}
\ee
where  $\Omega$ and  $\alpha$ are real constants. Such curved H\'enon--Heiles Hamiltonian will be constructed by considering it as an integrable cubic perturbation of the 1:2 anisotropic oscillator that we have introduced in the previous Section~\ref{sec52}, in the form (\ref{xzy}), although in this case projective Beltrami coordinates of Section~\ref{sec3} will be the ones that are naturally adapted to the construction of the curved system.  

We recall that the original (non-integrable) H\'enon--Heiles system 
$$
{H}=\dfrac{1}{2}(p_{1}^{2}+p_{2}^{2}) + \dfrac{1}{2}(q_{1}^{2}+q_{2}^{2})+\lambda\left(
q_{1}^{2}q_{2}-\frac{1}{3}\,q_{2}^{3}\right),
\label{HHaut}
$$
was introduced in~\cite{HH} in order to model a Newtonian axially-symmetric galactic system. When the following generalization containing adjustable parameters was studied
\begin{equation}
\mathcal{H}=\dfrac{1}{2}(p_{1}^{2}+p_{2}^{2})+ \Omega_{1}  q_{1}^{2}+\Omega_{2} q_{2}^{2}+\alpha \left(
q_{1}^{2}q_{2}+\beta q_{2}^{3}\right) ,
\label{hhmulti}\nonumber
\end{equation}
it was found that the only Liouville-integrable members of this family of generalized   H\'enon--Heiles Hamiltonians
were given by   {\em three} specific choices of the real parameters $\Omega_1$, 
$\Omega_2$, $\alpha$ and $\beta$ (see~\cite{BSV,CTW,GDP,HietarintaRapid,Fordy83,Wojc,SL,FordyHH,Sarlet,RGG,Tondo,Conte,HHjpcs}):

\begin{itemize} 

\item The Sawada--Kotera system, given by $\beta=1/3$ and $\Omega_{1}=\Omega_{2}=\Omega$:
\begin{equation}
 \mathcal{H}=\dfrac{1}{2}(p_{1}^{2}+p_{2}^{2})+ \Omega \left( q_{1}^{2}+ q_{2}^{2}\right)+\alpha \left(
q_{1}^{2}q_{2}+\frac 13 q_{2}^{3}\right) .
\label{HSK1}
 \end{equation}
 This system is separable in rotated Euclidean coordinates, and therefore its integral of the motion is quadratic in the momenta.\\
 
\item  The Korteweg--de Vries (KdV) system, with $\beta=2$ and $(\Omega_{1},\Omega_{2})$ arbitrary parameters:
\begin{equation}
 \mathcal{H}=\dfrac{1}{2}(p_{1}^{2}+p_{2}^{2})+\Omega_{1}  q_{1}^{2}+\Omega_{2} q_{2}^{2}+\alpha \left(
q_{1}^{2}q_{2}+2 q_{2}^{3}\right),
\label{HKdV1}
 \end{equation}
which is separable in parabolic coordinates and has also a quadratic integral of the motion.\\
 
\item The Kaup--Kuperschdmit system, with $\beta=16/3$ and $\Omega_{2}=16\Omega_{1} =16\Omega$:
\begin{equation}
 \mathcal{H}=\dfrac{1}{2}(p_{1}^{2}+p_{2}^{2})+ \Omega \left( q_{1}^{2}+ 16 q_{2}^{2}\right)+\alpha \left(
q_{1}^{2}q_{2}+\frac {16}3 q_{2}^{3}\right),
\label{HKK1}
 \end{equation}
whose integral is quartic in the momenta.

\end{itemize}

Hence the particular KdV case (\ref{HKdV1}) arising when $\Omega_{2}=4\Omega_{1}$ gives the Hamiltonian (\ref{KdVflat}) and this is
 connected to the so-called  {Ramani--Dorizzi--Grammaticos (RDG)  series} of integrable potentials~\cite{RDGprl,Hietarinta}, which are just the polynomial potentials on the Euclidean plane that can be separated in parabolic coordinates and can freely be  superposed  by preserving integrability~\cite{Wojc,FF}.  Moreover, such separability in parabolic coordinates explains why a large collection of integrable rational perturbations can be added to the RDG potentials (see~\cite{FF,HoneIP,HonePLA,tesis,Annals10} and references therein).

In   the sequel we  review the main results concerning the flat KdV H\'enon--Heiles    Hamiltonian (\ref{HKdV1}) with $\Omega_{2}=4\Omega_{1}$  along with its associated RDG potentials. And
in   the next  Section~\ref{sec7} we  will  sketch its integrable curved analogue on the 2D sphere \textbf{S}$^{2}$ and the   hyperbolic (or Lobachevski) space  \textbf{H}$^{2}$  which was constructed in~\cite{HHnon}, together with the full curved counterpart of the integrable RDG series of potentials. The corresponding integrable perturbations of the curved KdV system can be found in~\cite{HHjpcs}. 

%%%%%%%%%%%%%%%%%%%%%%%%%%%%%%%%%%%%%%%%%%%%%%%

\subsection{An integrable KdV H\'enon--Heiles system on the Euclidean plane}
\label{sec61}

Le us consider the   integrable (albeit non-superintegrable) Hamiltonian system (\ref{KdVflat}) defined on  \textbf{E}$^{2}$  whose constant of motion is quadratic in the momenta and given by
\begin{equation}
\mathcal{I} = 
p_{1}(q_{1}p_{2}-q_{2}p_{1})+q_{1}^{2}\left(2\Omega q_{2}+
\dfrac{\alpha}{4}(q_{1}^{2}+4 q_{2}^{2})
\right) .
 \label{I2KdVflat}
\end{equation}
This system  can be regarded as an integrable cubic perturbation of the 1:2 oscillator with frequencies $(\omega, 2\omega)$ once the  identification $\omega^{2}=2\Omega$ is performed (see Section~\ref{sec42}).

The potential functions included in both the Hamiltonian (\ref{KdVflat}) and its invariant (\ref{I2KdVflat}) are directly connected to the so-called  RDG series  of integrable potentials, which consists of the homogeneous polynomial potentials of degree $n$ given by~\cite{RDGprl,Hietarinta}
\begin{equation}
\mathcal{V}_{n}(q_1,q_2) =\sum\limits_{i=0}^{[\frac{n}{2}]}2^{n-2i}\dbinom{n-i}{i}q_{1}^{2i}q_{2}^{n-2i}
\, ,\qquad n=1,2,\dots
\nonumber
\end{equation}
Namely, the four members of this family read
\bea
&& \mathcal{V}_{1}(q_{1},q_{2})=2q_{2} ,\nonumber \\ 
&& \mathcal{V}_{2}(q_{1},q_{2})=q_{1}^2+ 4q_{2}^2 ,\nonumber \\
&& \mathcal{V}_{3}(q_{1},q_{2})=4q_{1}^2q_{2}+ 8q_{2}^3 ,\nonumber \\
&& \mathcal{V}_{4}(q_{1},q_{2})=q_{1}^4+ 12 q_{1}^2 q_{2}^2 + 16 q_{2}^4  .
\nonumber
\eea
It is straightforward to realize that the quadratic and cubic potentials in the Hamiltonian (\ref{KdVflat}) are just the 
 second- and the third-order RDG potentials  $\mathcal{V}_{2}$ and $\mathcal{V}_{3}$, respectively. Moreover, the integral  $\mathcal{I}$   (\ref{I2KdVflat}) contains the linear   $ \mathcal{V}_{1}$ and the  quadratic $ \mathcal{V}_{2}$ RDG potentials. Therefore, the integrable system~\eqref{KdVflat} is constructed through the building block functions $\mathcal{V}_{1}$, $\mathcal{V}_{2}$ and $\mathcal{V}_{3}$.
 
In fact, it can be straightforwardly proven that a Hamiltonian $\mathcal{H}_{n}$ containing the RDG potential  $\mathcal{V}_{n}$, namely,
 \begin{equation}
\mathcal{H}_{n}=\dfrac{1}{2}(p_{1}^{2}+p_{2}^{2})+\alpha_{n}\mathcal{V}_{n},
\nonumber
\end{equation}
is always Liouville integrable, with integral of the motion $\mathcal{L}_n$ involving  the $\mathcal{V}_{n-1}$ potential in the form
 \begin{equation}
\mathcal{L}_n=p_{1}(q_{1}p_{2}-q_{2}p_{1})+ \alpha_{n} q_{1}^{2}\mathcal{V}_{n-1} ,\quad\  \{\mathcal{H}_{n},\mathcal{L}_n\}=0 .
\label{Rflat}
\end{equation}
Note that  formula (\ref{Rflat}) holds provided that the 0-th order RDG potential is defined as the constant
$\mathcal{V}_{0}:=1$, and the first integrable Hamiltonian system within the RDG series reads
 \begin{equation}
\mathcal{H}_{1}=\dfrac{1}{2}(p_{1}^{2}+p_{2}^{2})+\alpha_{1}  ( 2 q_2  ), \qquad 
\mathcal{L}_1=p_{1}(q_{1}p_{2}-q_{2}p_{1})+ \alpha_{1} q_{1}^{2} .
\label{Rflat1}\nonumber
\end{equation}
Furthermore, all RDG potentials can freely be  superposed by preserving   integrability~\cite{Hietarinta,tesis,Annals10}. More explicitly, the Hamiltonian
\begin{eqnarray}
&& \mathcal{H}_{(M)}=\dfrac{1}{2}\left(
p_{1}^{2}+p_{2}^{2}
\right)+\sum\limits_{n=1}^M
\alpha_n \mathcal{V}_{n} \nonumber \\[2pt]
&&\qquad \ \,
=\dfrac{1}{2}\left(p_{1}^{2}+p_{2}^{2}\right)+\sum\limits_{n=1}^{M}\sum\limits_{i=0}^{[\frac{n}{2}]}\alpha_{n}2^{n-2i}\dbinom{n-i}{i}q_{1}^{2i}q_{2}^{n-2i} ,
\label{bk}
\end{eqnarray}
where $M=1,2,\dots$ and $\alpha_n$ are arbitrary real constants, 
has the following  integral of the motion
\begin{eqnarray}
&&\!\!\!\!\!\!\!\!\!\!\!\!\!\!\! \mathcal{L}_{(M)} =  p_{1}(q_{1}p_{2}-q_{2}p_{1})
+q_{1}^{2}\sum\limits_{n=1}^{M}\alpha_{n}\mathcal{V}_{n-1} \nonumber\\[2pt]
&&\!\!\!\!\ \,   = p_{1}(q_{1}p_{2}-q_{2}p_{1})
+q_{1}^{2}\left(
\sum\limits_{n=1}^{M}\sum\limits_{i=0}^{[\frac{n-1}{2}]}\alpha_{n}2^{n-1-2i}\dbinom{n-1-i}{i}q_{1}^{2i}q_{2}^{n-1-2i}\right)\! .
\label{I2Ramani}
\end{eqnarray}

Therefore, the KdV H\'enon--Heiles Hamiltonian $ \mathcal{H} $ (\ref{KdVflat}) and its integral $ \mathcal{I}$ (\ref{I2KdVflat}) can be thought of as the Hamiltonian $\mathcal{H}_{(M)} $  (\ref{bk}) and the integral  $\mathcal{L}_{(M)}$ (\ref{I2Ramani})  by setting
\begin{equation}
M=3,\qquad \alpha_1=0,\qquad 
\alpha_2=\Omega,\qquad  \alpha_3= \alpha/4 ,\label{const}
\end{equation}
since in that case we obtain that
\begin{eqnarray}
&&  \mathcal{H}_{(3)} =\dfrac{1}{2}(p_{1}^{2}+p_{2}^{2})+\alpha_{2} \mathcal{V}_{2}+\alpha_{3} \mathcal{V}_{3} ,
\nonumber\\
&& \mathcal{L}_{(3)}=  
p_{1}(q_{1}p_{2}-q_{2}p_{1})+q_{1}^{2}\left(\alpha_2\mathcal{V}_{1} +
 {\alpha_3} \mathcal{V}_{2}   
\right)  .
\label{be}\nonumber
\end{eqnarray}
As we will see in the sequel, this integrability structure associated to the RDG potentials can be fully generalized after introducing the integrable deformation generated by the curvature parameter.

%%%%%%%%%%%%%%%%%%%%%%%%%%%%%%%%%%%%%%%%%%%%%%%

\section{An integrable KdV H\'enon--Heiles system  on ${\>S}^{2}$ and ${\>H}^{2}$ }
\label{sec7}

The curved counterpart of the KdV H\'enon--Heiles system (\ref{KdVflat}) was constructed in~\cite{HHnon} by making use of the approach we advocate in this paper, which can be summarized as follows. Given an integrable Euclidean H\'enon--Heiles system
\be
\mathcal{H}=\mathcal{T} + \mathcal{V}=\dfrac{1}{2}(p_{1}^{2}+p_{2}^{2}) + \mathcal{V}_2(q_1,q_2)+ \mathcal{V}_3(q_1,q_2),
\nonumber
\ee
an integrable generalization of this system to \textbf{S}$^{2}$ and \textbf{H}$^{2}$ of the form
\be
\mathcal{H}_\kappa=\mathcal{T}_\kappa(p_1,p_2,q_1,q_2) + \mathcal{V}_{\kappa,2}(q_1,q_2)+ \mathcal{V}_{\kappa,3}(q_1,q_2),
\label{nu123}
\ee
is constructed through the following steps:

\begin{enumerate}

\item Use the projective coordinates presented in Section~\ref{sec3} in order to describe the free motion on \textbf{S}$^{2}$ and \textbf{H}$^{2}$ (so such kinetic energy  term $\mathcal{T}_\kappa$ is known  and given by (\ref{dc})).\\

\item  Take the integrable curved anisotropic 1:2 oscillator and its integral of the motion given in Section~\ref{sec52} in the form (\ref{xzy}) as the initial data in order to construct the  cuved family of RDG potentials.\\

\item Construct the full family of integrable curved RDG potentials on \textbf{S}$^{2}$ and \textbf{H}$^{2}$ (that we shall denote as $\mathcal{V}_{\k, n}$)  through a recurrence procedure.\\

\item Show that the curved RDG potentials can be superposed by preserving integrability.\\

\item Obtain the curved 1:2 KdV H\'enon--Heiles system as the particular case~\eqref{nu123} of the latter curved RDG system.
\end{enumerate}

Two important comments concerning this approach have to be pointed out: firstly, that projective coordinates will be the suitable ones in order to construct the curved RDG potentials and, secondly, that the integrability properties of $\mathcal{V}_{\kappa,2}$ will be our ``initial conditions" that will guide the construction of the full integrability structure.

By following this procedure (see~\cite{HHnon} for details), the  RDG potentials on the sphere  ${\>S}^{2}$ and   the  hyperbolic space  ${\>H}^{2}$ can be defined in terms of projective Beltrami  coordinates $(q_1,q_2)$ as
\begin{eqnarray}
&&\!\!\!\!\!\! \mathcal{V}_{\k, n}  =\left(\dfrac{1+\kappa   {\>q}^{2}}{1-\k q_{2}^{2}}\right)^{\! 2} \times 
\nonumber\\[2pt]
&&\qquad\qquad  \sum\limits_{i=0}^{[\frac{n}{2}]}2^{n-2i}\dbinom{n-i}{i}  \! \left(
\dfrac{q_{1}}{\sqrt{1+\kappa   {\>q}^{2}}}
\right)^{\! 2i} \!\! \left(
1-\dfrac{i }{n-i}\left[\dfrac{\kappa  q_{1}^{2}}{1+\kappa   {\>q}^{2}}\right]
\right)
\!\left(
\dfrac{q_{2}}{1+\kappa   {\>q}^{2}}
\right)^{\! n-2i}   
\nonumber
\end{eqnarray}
 with  $n=1,2,\dots$. It is straightforward to prove that each curved RDG Hamiltonian 
 \begin{equation}
 \mathcal{H}_{\k,n}  = {\cal T}_\k   +\alpha_{n} \mathcal{V}_{\k,n} ,\nonumber
 \end{equation}
is integrable, with integral of motion $\mathcal{L}_{\k,n}$ being quadratic in the momenta and given by
 \begin{equation}
\mathcal{L}_{\k,n} =J_{01}J_{12}+\alpha_{n} \,\dfrac{q_{1}^{2}}{1+ \kappa \bf{q}^{2} } \mathcal{V}_{\k, n-1} , \qquad \{ \mathcal{H}_{\k,n} ,\mathcal{L}_{\k,n}\}=0 ,
\nonumber
\end{equation}
 where   ${\cal T}_\k$ is the kinetic energy (\ref{dc}) and  $J_{01}$,  $J_{12}$      are the functions   given in Table~\ref{table1} in Beltrami variables. We stress that   in order to get a suitable recurrence relation, the 0-term $\mathcal{V}_{\k, 0} $ in the curved RGD series of potentials is by no means a constant and it has to be defined as the function
 \begin{equation}
  \mathcal{V}_{\k,0}:=\dfrac{(1+ \kappa q_{2}^{2})(1+ \kappa \>q^2  )}{\left(1- \kappa q_{2}^{2}\right)^{2}} \, .
\nonumber
\end{equation}
 
Note that   the quadratic curved RDG Hamiltonian,  $ \mathcal{H}_{\k,2}  = {\cal T}_\k   +\alpha_{2} \mathcal{V}_{\k,2} $,  is just the superintegrable curved 1:2 oscillator  (\ref{xzy}), formerly   introduced in~\cite{RS} and further studied in~\cite{BaHeMu13,BaBlHeMu14}.

It is convenient to recall that in terms of the  ambient coordinates $(x_0,x_1,x_2)$, subjected to the constraint (\ref{ci}), the first curved RDG potentials turn out to be
    \begin{eqnarray}
&&\!\!\!\!\!\! \mathcal{V}_{\k, 0}  =  \dfrac{1- \kappa x_{1}^{2}}{(x_{0}^{2}-\kappa x_{2}^{2})^{2}} , \nonumber \\[2pt]   
&&\!\!\!\!\!\! {V}_{\k,1} =\dfrac{2 x_0x_2  }{(x_{0}^{2}-\kappa x_{2}^{2})^{2}} \, , \nonumber \\[2pt] 
&&\!\!\!\!\!\!  {V}_{\k,2}=   \frac{x_1^2 (1-\k x_1^2)+  4 x_0^2x_2^2  }{(x_{0}^{2}-\kappa x_{2}^{2})^{2}} \, , \nonumber \\[2pt] 
&&\!\!\!\!\!\! {V}_{\k,3}=   \dfrac{ 4 x_0 x_1^2  x_2 (1-\frac 12 \k x_1^2)+ 8  x_0^3x_2^3  }{(x_{0}^{2}-\kappa x_{2}^{2})^{2}} \, , 
\nonumber
\end{eqnarray}
and the general formula for the curved RGD potentials is given by
  \begin{equation}
\mathcal{V}_{\k, n}   =
\dfrac{1}{(x_{0}^{2}-\kappa x_{2}^{2})^{2}}\sum\limits_{i=0}^{[\frac{n}{2}]}2^{n-2i}\dbinom{n-i}{i}x_{1}^{2i}\left(
1-\dfrac{i}{n-i}\kappa  x_{1}^{2}
\right) \left(x_{0} x_{2}\right)^{n-2i} \, .
\nonumber
\end{equation}
Obviously, from these expressions these potentials can be written in any other coordinate system.   Notice  also that $ {V}_{\k,2}$ is exactly  the potential written in  (\ref{xzy}) due to the relation  (\ref{ci}).

As in the Euclidean case the curved RDG potentials can be superposed and therefore expressions (\ref{bk}) and (\ref{I2Ramani})  can be generalized to the curved case~\cite{HHnon}.
In this way, it can  straightforwardly be shown that the Hamiltonian 
\begin{equation}
\mathcal{H}_{\kappa,(M)}= {\cal T}_\k +\sum\limits_{n=1}^{M}\alpha_{n}\mathcal{V}_{\k,n} \, , \qquad M=1,2,\dots  
\nonumber
 \end{equation}
Poisson-commutes with  the function
\begin{equation}
\mathcal{L}_{\kappa,(M)}= J_{01}J_{12}+\dfrac{q_{1}^{2}}{1+ \kappa  \bf{q}^{2}}    \sum\limits_{n=1}^{M}\alpha_{n}\mathcal{V}_{\k, n-1} \, , 
\nonumber
\end{equation}
where $J_{01}$,  $J_{12}$  and  ${\cal T}_\k$  are again given in Table~\ref{table1}.

Finally, the integrable curved counterpart of the H\'enon--Heiles KdV Hamiltonian~\eqref{KdVflat} on ${\mathbf S}^2$ and ${\mathbf H}^2$ arises as a straightforward corollary of the previous result as the particular case $\mathcal{H}_{\kappa,(3)}$ and by considering (\ref{const}). Explicitly,
    \begin{equation}
\mathcal{H}_\kappa  ={\cal T}_\k + {\cal V}_\k={\cal T}_\k +\Omega\, \mathcal{V}_{\kappa, 2}+\frac {\alpha}4\, \mathcal{V}_{\kappa, 3} \, ,
\nonumber
\end{equation}
and the curved analogue of the H\'enon--Heiles KdV potential is so given by
    \begin{equation}
{\cal V}_\k= \Om\,   \frac{  q_1^2(1+\k q_2^2) + 4 q_2^2}{(1-\kappa  \tq_2^2)^2  }  +
 {\alpha}\,  \frac{   q_1^2 q_2(1+\k\>q^2- \frac 12 \k q_1^2)  + 2q_2^3}{ (1-\kappa  \tq_2^2)^2 (1+\k \>q^2)  }  \, .
\nonumber
\end{equation}
The associated integral of the motion  comes from $\mathcal{L}_{\kappa,(3)}$ and reads
 \begin{eqnarray}
&&  \mathcal{I}_{\kappa}=  J_{01}J_{12}+\frac{q_{1}^{2}}{1+ \kappa  \bf{q}^{2}} \left(\Omega\,  \mathcal{V}_{\k,1} +
\frac {\alpha}4\, \mathcal{V}_{\k,2}   
\right)\nonumber\\[2pt]
&&\quad\ =   \left(   p_1+\kappa (\bq\cdot\bp) q_1  \right)\left( q_1 p_2 - q_2 p_1\right) \nonumber \\[2pt]
&& \qquad\qquad+\frac{q_{1}^{2}}{1+ \kappa  \bf{q}^{2}} \left(\Omega\, \frac{2 q_{2}(1+\kappa  {\>q}^{2})}{(1-\kappa  q_{2}^{2})^{2}}+
 {\alpha} \, \frac{  q_1^2(1+\k q_2^2) + 4 q_2^2}{4(1-\kappa  \tq_2^2)^2  }  
\right)\, .
 \nonumber
\end{eqnarray}
We stress that, by construction, the $\kappa\to 0$ limit of all these expressions leads smoothly to their Euclidean counterparts (\ref{KdVflat}) and (\ref{I2KdVflat}) we started with.

%%%%%%%%%%%%%%%%%%%%%%%%%%%%%%%%%%%%%%%%%%%%%%%%%%%%%%%%%%%%

\section{Remarks and open problems}

In this contribution we have intended to provide a summary of recent results concerning the construction of new (super)integrable systems on 2D spaces of constant curvature as (super)integrable deformations of the corresponding Euclidean systems, where the Gaussian curvature of the space plays the role of the parameter for an integrable deformation theory.

This approach can be developed in different coordinate systems, and we have stressed the fact that
projective Beltrami coordinates are computationally very useful from the viewpoint of algebraic integrability, since in these coordinates the curved kinetic energy is just a polynomial in the canonical projective variables and the curved integrable potentials so obtained can be expressed as rational functions. As a summarizing example illustrating this fact we recall that the Higgs oscillator Hamiltonian ${\cal H}_\k^{1:1}$~\cite{Higgs,Serret} (this is just the 1:1 oscillator on ${\>S}^{2}$ and ${\>H}^{2}$ presented in Section~\ref{sec51}) is expressed, respectively, in terms of ambient, geodesic polar and Beltrami canonical variables as follows:
  \bea
  &&{\cal H}_\k^{1:1}=  \frac{\kk\left(x_1   \pp_1 + x_2   \pp_2 \right)^2}{2\left(1-  \kk \left(x_1^2+   x_2^2\right)\right)}+  \frac 12 \bigl(  \pp_1^2+      \pp_2^2\bigl)+ \,  \del \,\frac{\>x^2}{(1-\k \>x^2)}  \, ,
  \nonumber\\
  && \quad\quad\  =  \frac 12\left(p_r^2+  \frac{p_\phi^2}{ \Sk_{\kk}^2(r)}   \right)+ \del \Tk^2_\kk(r)  \, ,\nonumber
  \\
   &&\quad\quad\  =  \tfrac 12  (1+\kk\,\>q^2)\bigl( \>p^2+ \kk (\>q \cdot {   {\>p} } )^2 \bigr)+  \del\, \bq^2 \, .
\nonumber
  \eea
The computational advantages of the projective dynamics approach become evident from these expressions, specially for the search of curved analogues of non-superintegrable systems (like H\'enon--Heiles ones) where the lack of additional symmetries implies the need of making use of  a purely computational approach. We also recall that in terms of Beltrami coordinates the superintegrable Kepler--Coulomb potential on ${\>S}^{2}$ and ${\>H}^{2}$ is given by ${\cal V}^{{\rm KC}}=k/\sqrt{\bq^2}$ (see~\cite{LetterBH,Kepler}), where again the potential in projective coordinates coincides formally with its corresponding Euclidean expression, and all the dynamical modifications arising from a non-vanishing curvature are concentrated in the kinetic energy term.

It should also be  stressed that both anisotropic Euclidean oscillators and the integrable H\'enon--Heiles Hamiltonian here considered preserve their integrability under the addition of some centrifugal terms, and the curved analogues of these ``centrifugally perturbed" Hamiltonians can also be  constructed. On the other hand, the wide applicability of the method here presented is currently being used in order to construct the curved analogue of the KdV H\'enon--Heiles system (\ref{HKdV1}) for arbitrary $\Omega_1$ and $\Omega_2$ parameters, as well as the curved analogue of the Sawada--Kotera case (\ref{HSK1})  as an integrable curvature perturbation of the Higgs 1:1 oscillator. Also, the construction of the curved Kaup--Kuperschdmit H\'enon--Heiles Hamiltonian (\ref{HKK1}) should be based on the constant curvature analogue of the superintegrable 1:4 curved oscillator, and is currently under investigation. 

Finally, two further generalizations of the approach here presented should be mentioned. The first of them is the construction of integrable curved analogues of Minkowskian (instead of Euclidean) integrable systems, which could be addressed by following the same curvature-deformation approach, but considering the corresponding relativistic geometries with constant curvature (see \cite{CK2,trigo,conf,kiev,Petrosian1} and references therein). The second one deals with the construction of integrable systems on spaces with non-constant curvature, which in some cases can also be  considered as (quantum) deformations of known (super)integrable systems on the Euclidean space. In these cases, a quite similar approach based on integrable perturbations in terms of a parameter related with the curvature has led to the obtention of new superintegrable oscillator and Kepler--Coulomb potentials on Darboux III and Taub-NUT spaces  (see~\cite{KaKrWint02,intQG,PhysD2008,Ballestetal08ClQGr,ComRunge,WHS,RagRig10,Sigma11,BurgosAnnPh11,taub} for further details and references on integrability on spaces with non-constant curvature).

%%%%%%%%%%%%%%%%%%%%%%%%%%%%%%%%%%%%%%

\section*{Acknowledgements}

This work has been partially supported by Ministerio de Ciencia, Innovaci\'on y Universidades (Spain) under grant MTM2016-79639-P (AEI/FEDER, UE) and by Junta de Castilla y Le\'on (Spain) under grant BU229P18.

%%%%%%%%%%%%%%%%%%%%%%%%%%%%%%%%%%%%%%


\newpage
 
\begin{thebibliography}{99}
\small


\bibitem{Perelomov}
Perelomov A.M.:  Integrable systems of classical mechanics and Lie algebras. 
 Berlin, Birkh\"auser (1990)
 
\bibitem{Goriely}
Goriely  A.:  Integrability and nonintegrability of dynamical systems. 
 Singapore, World Scientific (2001)
 

  
 \bibitem{Vozmischeva}
Vozmischeva T.G.:   Integrable problems of celestial mechanics in spaces of constant curvature.
  Astrophysics and Space Science Library, vol.~295. Kluwer,  Dordrecht (2003)
  
   \bibitem{BoPu} Boccaletti  D.,    Pucacco  G.:  Theory of Orbits. Berlin, Springer (2004).


  \bibitem{MiPWJPa13}    Miller W.Jr.,   Post  S.,    Winternitz P.: 
Classical and quantum superintegrability with applications.    
J. Phys. A: Math. Theor.  {\bf  46},   423001  (2013) \href{https://doi.org/10.1088/1751-8113/46/42/423001}{\path{doi:10.1088/1751-8113/46/42/423001}}.     

\bibitem{CK2}
 Ballesteros A.,  Herranz F.J.,  del Olmo  M.A.,  Santander  M.: 
Quantum structure of the motion groups of the two-dimensional Cayley--Klein geometries.
{J. Phys.  A: Math. Gen.}  {\bf 26},  5801--5823 (1993) \href{https://doi.org/10.1088/0305-4470/26/21/019}{\path{doi:10.1088/0305-4470/26/21/019}}.


 \bibitem{RS}  Ra\~nada M.F.,  Santander M.:
Superintegrable systems on the two-dimensional sphere $S^2$ and the 
hyperbolic plane $H^2$. 
J. Math. Phys.  {\bf  40},    5026--5057  (1999) \href{https://doi.org/10.1063/1.533014}{\path{doi:10.1063/1.533014}}


\bibitem{trigo}
 Herranz  F.J.,  Ortega R.,  Santander M.:   Trigonometry of spacetimes: A new self-dual approach to a
      curvature/signature (in)dependent trigonometry.
 {J. Phys.  A: Math. Gen.} {\bf 33},   4525--4551 (2000) \href{https://doi.org/10.1088/0305-4470/33/24/309}{\path{doi:10.1088/0305-4470/33/24/309}}
 


\bibitem{conf}
 Herranz F.J.,  Santander M.:   Conformal symmetries of spacetimes.
{J. Phys. A: Math. Gen.} {\bf 35},   6601--6618 (2002) \href{https://doi.org/10.1088/0305-4470/35/31/306}{\path{doi:10.1088/0305-4470/35/31/306}}.

  \bibitem{ran}
    Ra\~nada  M.F.,    Santander M.:    On some properties of harmonic oscillator on spaces
of constant curvature. {Rep. Math. Phys.}  {\bf 49},   335--343  (2002) \href{https://doi.org/10.1016/S0034-4877(02)80031-3}{\path{doi:10.1016/S0034-4877(02)80031-3}}.

 \bibitem{ran1}
    Ra\~nada  M.F.,    Santander M.:    On harmonic oscillators on the two-dimensional sphere $S^2$ and
the hyperbolic plane $H^2$. {J. Math. Phys.} {\bf 43},  431--451 (2002) \href{https://doi.org/10.1063/1.1423402}{\path{doi:10.1063/1.1423402}}.

\bibitem{BaHeSantS03}  Ballesteros A.,    Herranz F.J.,   Santander M.,    Sanz-Gil T.:
Maximal superintegrability on $N$-dimensional curved spaces. 
J. Phys. A: Math. Gen.   {\bf  36},    L93--L99  (2003) \href{https://doi.org/10.1088/0305-4470/36/7/101}{\path{doi:10.1088/0305-4470/36/7/101}}.  

\bibitem{CRMVulpi}
 Herranz F.J., Ballesteros A., Santander M., Sanz-Gil  T.:   Maximally superintegrable {S}morodinsky--{W}internitz systems on the {N}-dimensional sphere and hyperbolic spaces.   {S}uperintegrability in {C}lassical and {Q}uantum {S}ystems,  {CRM} {P}roceedings and {L}ecture {N}otes, vol.~37, pp. 75--89,  ed. P.~Tempesta  et al.  American Mathematical Society, Providence, R.I., (2004) \href{https://doi.org/10.1090/crmp/037}{\path{doi:10.1090/crmp/037}}.

 \bibitem{kiev}  Herranz F.J., Ballesteros A.:   Superintegrability on three-dimensional {R}iemannian and relativistic spaces
of constant curvature.   {SIGMA}  Symmetry Integrability Geom. Methods Appl. {\bf 2}.  010 (2006) \href{https://doi.org/10.3842/SIGMA.2006.010}{\path{doi:10.3842/SIGMA.2006.010}}.

\bibitem{LetterBH}   Ballesteros A.,     Herranz  F.J.:
Universal integrals for superintegrable systems on $N$-dimensional spaces of constant curvature.  
J. Phys. A: Math. Theor. {\bf 40},   F51--F59  (2007) \href{https://doi.org/10.1088/1751-8113/40/2/F01}{\path{doi:10.1088/1751-8113/40/2/F01}}.

\bibitem{Santander6}
Cari\~nena J.F., Ra\~nada M.F., Santander M.: The quantum harmonic oscillator on the sphere and the hyperbolic plane. {Ann. Phys.} {\bf 322},   2249--2278 (2007) \href{https://doi.org/10.1016/j.aop.2006.10.010}{\path{doi:10.1016/j.aop.2006.10.010}}.

\bibitem{CRS07JPa}   Cari\~nena J.F.,  Ra\~nada M.F.,  Santander M.:
Superintegrability on curved spaces, orbits and momentum hodographs: revisiting a classical result by Hamilton,   J. Phys. A: Math. Theor. {\bf 40},   13645--13666   (2007) \href{https://doi.org/10.1088/1751-8113/40/45/010}{\path{doi:10.1088/1751-8113/40/45/010}}. 

\bibitem{CRS08}    Cari\~nena J.F.,    Ra\~nada M.F.,   Santander M.:  
The {K}epler problem and the {L}aplace-{R}unge-{L}enz vector on spaces of constant curvature and arbitrary signature.  
 Qual. Theory Dyn. Syst.  {\bf   7},  87--99  (2008) \href{https://doi.org/10.1007/s12346-008-0004-3}{\path{doi:10.1007/s12346-008-0004-3}}. 

\bibitem{Kepler}
Ballesteros A., Herranz F. J.:  Maximal superintegrability of the generalized
{K}epler--{C}oulomb system on N-dimensional
curved spaces.
{J. Phys. A: Math. Theor.}
{\bf  42}, 245203  (2009)  \href{https://doi.org/10.1088/1751-8113/42/24/245203}{\path{doi:10.1088/1751-8113/42/24/245203}}.
   

\bibitem{DiacuJDE} Diacu F.,  P\'erez-Chavela E.: Homographic solutions of the curved  3-body problem {J. Diff. Equations} {\bf 250}, 340--366 (2011) \href{https://doi.org/10.1016/j.jde.2010.08.011}{\path{doi:10.1016/j.jde.2010.08.011}}.

\bibitem{Diacu1} Diacu  F.,  P\'erez-Chavela  E., Santoprete  M.:  The $n$-body problem in spaces of constant curvature. Part I: Relative Equilibria. 
{J. Nonlinear Sci.} {\bf 22}, 247--266 (2012) \href{https://doi.org/10.1007/s00332-011-9116-z}{\path{doi:10.1007/s00332-011-9116-z}}.


\bibitem{Diacu2} Diacu  F.,  P\'erez-Chavela  E., Santoprete  M.: The $n$-body problem in spaces of constant curvature. Part II: Singularities Equilibria. 
{J. Nonlinear Sci.} {\bf 22}, 267--275 (2012) \href{https://doi.org/10.1007/s00332-011-9117-y}{\path{doi:10.1007/s00332-011-9117-y}}.


\bibitem{DiacuM} Diacu F.:   Relative equilibria in the 3-dimensional curved $n$-body problem. {Memoirs Amer. Math. Soc.} {\bf 228}, 1071 (2014)  \href{http://dx.doi.org/10.1090/memo/1071}{\path{doi:10.1090/memo/1071}}.



\bibitem{GonKas14AnnPhys}     Gonera C.,    Kaszubska   M.: 
Superintegrable systems on spaces of constant curvature.  
Ann. Phys.  {\bf  364},  91--102  (2014) \href{https://doi.org/10.1016/j.aop.2014.04.005}{\path{doi:10.1016/j.aop.2014.04.005}}.   

\bibitem{Ra14JPaTTWk}   Ra\~nada M.F.:
The {T}remblay-{T}urbiner-{W}internitz  system on spherical and hyperbolic spaces : Superintegrability, curvature-dependent formalism and complex factorization. 
J. Phys. A: Math. Theor.  {\bf 47}, 165203  (2014) \href{https://doi.org/10.1088/1751-8113/47/16/165203}{\path{doi:10.1088/1751-8113/47/16/165203}}.   

\bibitem{Mfran15PLa}  Ra\~nada  M.F.:
The {P}ost-{W}internitz system on spherical and hyperbolic spaces: a proof of the superintegrability making use of complex functions and a curvature-dependent formalism.    
Phys. Lett. A {\bf 379},   2267--2271   (2015) \href{https://doi.org/10.1016/j.physleta.2015.07.043}{\path{doi:10.1016/j.physleta.2015.07.043}}.   
 
\bibitem{Ra15Jmp}    Ra\~nada   M.F.:
Superintegrable deformations of superintegrable systems: {Q}uadratic superintegrability and higher-order superintegrability.
J. Math. Phys. {\bf 56},   042703 (2015) \href{https://doi.org/10.1063/1.4918611}{\path{doi:10.1063/1.4918611}}. 

 
 \bibitem{Chanu}   Chanu  C.M.,   Degiovanni L.,   Rastelli   G.:
Warped product of Hamiltonians and extensions of Hamiltonian systems.    
 J.  Phys.: Conf. Ser.  {\bf  597},   012024  (2015) \href{https://doi.org/10.1088/1742-6596/597/1/012024}
 {\path{doi:10.1088/1742-6596/597/1/012024}}


 
\bibitem{Albouy2013}
 Albouy A.: There is a projective dynamics.  {Eur. Math. Soc. Newsl. } {\bf 89}, 
 37--43 (2013) \href{http://www.ems-ph.org/journals/newsletter/pdf/2013-09-89.pdf}{\path{http://www.ems-ph.org/journals/newsletter/pdf/2013-09-89.pdf}}.

\bibitem{Albouy2015} 
 Albouy A.: Projective dynamics and first integrals.  {Regul. Chaot. Dyn. } {\bf 20}, 247--276 (2015)
 \href{https://doi.org/10.1134/S1560354715030041}{\path{doi:10.1134/S1560354715030041}}.

 
 \bibitem{BaHeMu13}    Ballesteros  A.,     Herranz F.J.,  Musso F.:
The anisotropic oscillator on the 2D sphere and the hyperbolic plane. 
Nonlinearity   {\bf  26},    971--990  (2013) \href{http://iopscience.iop.org/article/10.1088/0951-7715/26/4/971/pdf}{\path{doi: 10.1088/0951-7715/26/4/971}}.  

\bibitem{BaBlHeMu14}   Ballesteros A.,     Blasco  A.,    Herranz  F.J.,  Musso  F.:
A new integrable anisotropic oscillator on the two-dimensional sphere and the hyperbolic plane.
J. Phys. A: Math. Theor.  {\bf 47},     345204  (2014)   \href{http://iopscience.iop.org/article/10.1088/1751-8113/47/34/345204/pdf}{\path{doi: 10.1088/1751-8113/47/34/345204}}.

 \bibitem{IW}
In\"on\"u E.,  Wigner  E.P.:
On the contractions of groups and their representations.
 {Proc. Natl. Acad. Sci. U.S.A.}
{\bf  39}, 510--524 (1953) \href{https://doi.org/10.1073/pnas.39.6.510}{\path{doi:10.1073/pnas.39.6.510}}.


\bibitem{Montigny}
Herranz  F.J.,   de Montigny  M.,  del Olmo  M.A.,   Santander  M.: 
{C}ayley--{K}lein algebras as graded contractions of $so(N+1)$.
{J. Phys.  A: Math. Gen.}  {\bf 27},  2515--2526 (1994) \href{https://doi.org/10.1088/0305-4470/27/7/027}{\path{doi:10.1088/0305-4470/27/7/027}}.


\bibitem{Yaglom}
Yaglom  I.M.: A Simple {N}on-{E}uclidean {G}eometry and its {P}hysical {B}asis.  
Springer, New York  (1979)


\bibitem{Groma}
Gromov N.A.,  Man'ko V.I.: 
 The {J}ordan--{S}chwinger representations of
{C}ayley--{K}lein groups. I. The orthogonal groups.
{J. Math. Phys.} {\bf 31}   1047--1053 (1990) \href{https://doi.org/10.1063/1.528781}{\path{doi:10.1063/1.528781}}.

\bibitem{Doubrovine}
Doubrovine B., Novikov  S.,  Fomenko A.:   G\'eom\'etrie Contemporaine, M\'ethodes et Applications First
Part. Moscow, MIR (1982).

  
 \bibitem{Jauch}  Jauch  J.M.,    Hill  E.L.:   On the problem of degeneracy in quantum mechanics. {Phys. Rev.} {\bf  57}, 641--645 (1940) \href{https://doi.org/10.1103/PhysRev.57.641}{\path{doi:10.1103/PhysRev.57.641}}.

 \bibitem{Stefan}  
Amiet J.P.,  Weigert S.: Commensurate harmonic oscillators: {C}lassical symmetries. {J. Math. Phys.}  {\bf 43},  4110--4126 (2002)
\href{https://doi.org/10.1063/1.1488672}{\path{doi:10.1063/1.1488672}}.


\bibitem{Tempesta}
Rodr\1guez M.A.,      Tempesta P.,  Winternitz P.: Reduction of superintegrable systems: The anisotropic harmonic oscillator.  {Phys. Rev. E} {\bf 78}, 046608  (2008) \href{https://doi.org/10.1103/PhysRevE.78.046608}{\path{doi:10.1103/PhysRevE.78.046608}}.

 
\bibitem{Kuruannals}
  Ballesteros  A.,  Herranz F.J., Kuru  S., Negro J.: The anisotropic oscillator on curved spaces: A new exactly solvable model. {Ann.~Phys.} {\bf 373}, 399--423 (2016)  \href{https://doi.org/10.1016/j.aop.2016.07.006}{\path{doi:10.1016/j.aop.2016.07.006}}.

    
   \bibitem{KuruProcs}
  Ballesteros A.,  Herranz  F.J., Kuru S., Negro J.: Factorization approach to superintegrable systems: Formalism and applications. {Phys. Atom. Nuclei} {\bf 80}, 389--396 (2017)  \href{https://doi.org/10.1134/S1063778817020053}{\path{doi:10.1134/S1063778817020053}}.

\bibitem{david}
Fern\'andez C.~D.J., Negro J.,  del Olmo M.A.: Group approach to the factorization of the radial oscillator equation.
{Ann.~Phys.}  {\bf 252},  386--412 (1996) \href{https://doi.org/10.1006/aphy.1996.0138}{\path{doi:10.1006/aphy.1996.0138}}.

  
\bibitem {Kuru1} 
Kuru S., Negro J.:  Factorizations of one-dimensional classical systems. {Ann. Phys.} {\bf 323},   413--431 (2008) \href{https://doi.org/10.1016/j.aop.2007.10.004}{\path{doi:10.1016/j.aop.2007.10.004}}.

  
\bibitem{kuru12} Calzada  J.A., Kuru S., Negro J., del Olmo M.A.:  Dynamical algebras of general two-parametric {P}\"{o}schl--{T}eller {H}amiltonian. {Ann. Phys.} {\bf 327},  808--822 (2012) \href{https://doi.org/10.1016/j.aop.2011.12.014}{\path{doi:10.1016/j.aop.2011.12.014}}.

 
\bibitem{Kuru3}
Celeghini  E., Kuru S., Negro J.,  del Olmo M.A.: A unified approach to quantum and classical {TTW} systems based on factorizations.
{Ann.~Phys.}  {\bf 332},  27--37 (2013) \href{https://doi.org/10.1016/j.aop.2013.01.008}{\path{doi:10.1016/j.aop.2013.01.008}}.
 
 
 \bibitem{lissajous}
Calzada J.A., Kuru S.,    Negro J.: Superintegrable {L}issajous systems on the sphere.
{Eur. Phys. J. Plus}  {\bf 129},  129--164  (2014) \href{https://doi.org/10.1140/epjp/i2014-14164-5}{\path{doi:10.1140/epjp/i2014-14164-5}}.

 
 \bibitem{VerrierOscillator}
Evans  N.W., Verrier P.E.: Superintegrability of the caged anisotropic oscillator.      {J. Math. Phys.} {\bf 49},  092902  (2008) \href{https://doi.org/10.1063/1.2988133}{\path{doi:10.1063/1.2988133}}.


  \bibitem{Ev90Pra}  Evans N.W.:  
 Superintegrability in classical mechanics.
Phys. Rev.  A  {\bf  41},    5666--5676  (1990)  \href{https://doi.org/10.1103/PhysRevA.41.5666}{\path{doi:10.1103/PhysRevA.41.5666}}.


\bibitem{KaWiMiPo99}    Kalnins E.G.,   Williams G.C.,   Miller  W.,   Pogosyan  G.S.:
Superintegrability in the three--dimensional {E}uclidean space.
J. Math. Phys.  {\bf  40},     708--725  (1999)     \href{https://doi.org/10.1063/1.532699}{\path{doi:10.1063/1.532699}}.
  
  
 \bibitem{Demkov}  Demkov Yu. N.: 
Symmetry group of the isotropic oscillator. Soviet Phys. JETP {\bf 36},   63--66, (1959) \href{http://www.jetp.ac.ru/cgi-bin/dn/e_009_01_0063.pdf}{\path{http://www.jetp.ac.ru/cgi-bin/dn/e_009_01_0063.pdf}}.

\bibitem{Fradkin}   Fradkin D.M.:
Three-dimensional isotropic harmonic oscillator and $SU_3$. 
Amer. J. Phys. {\bf 33}, 207--211 (1965) \href{https://doi.org/10.1119/1.1971373}{\path{doi:10.1119/1.1971373}}.




\bibitem{FrMaSmUW65}   Fris T.I.,    Mandrosov V.,  Smorodinsky  Y.A.,   Uhlir M.,  Winternitz P.:
On higher symmetries in quantum mechanics.
Phys. Lett.  {\bf 16},  354--356  (1965) \href{https://doi.org/10.1016/0031-9163(65)90885-1}{\path{doi:10.1016/0031-9163(65)90885-1}}.



 \bibitem{Evans90}     Evans  N.W.:
Super-integrability of the {W}internitz system.  
Phys. Lett. A {\bf 147},   483--486 (1990) \href{https://doi.org/10.1016/0375-9601(90)90611-Q}{\path{doi:10.1016/0375-9601(90)90611-Q}}.


\bibitem{Evans91}     Evans  N.W.:
Group theory of the {S}morodinsky-{W}internitz system. 
J. Math. Phys. {\bf 32}, 3369--3375 (1991) \href{https://doi.org/10.1063/1.529449}{\path{doi:10.1063/1.529449}}.



\bibitem{GrPoSi95a}  Grosche C.,   Pogosyan G.S.,    Sissakian A.N.: 
Path integral discussion for {S}moro\-dinsky--{W}internitz potentials 
I. Two- and three dimensional Euclidean spaces.
Fortschr. Phys.  {\bf  43},  453--521  (1995)
 \href{https://doi.org/10.1002/prop.2190430602}{\path{doi:10.1002/prop.2190430602}}.
 
  

\bibitem{WolfBoyer} Wolf K.B., Boyer C.P.: 
The 2:1 anisotropic oscillator, separation of variables and symmetry group in {B}argmann space. {J. Math. Phys.} {\bf 16},   2215--2223 (1975) \href{https://doi.org/10.1063/1.522471}{\path{doi:10.1063/1.522471}}.

 
\bibitem{Higgs}   Higgs  P.W.: Dynamical symmetries in a spherical geometry I. {J. Phys. A: Math. Gen.} {\bf  12}, 309--323 (1979) \href{https://doi.org/10.1088/0305-4470/12/3/006}{\path{doi:10.1088/0305-4470/12/3/006}}.


 \bibitem{ran2}
    Ra\~nada  M.F.,    Santander M.:    On harmonic oscillators on the two-dimensional sphere $S^2$ and the hyperbolic plane $H^2$ II.
  {J. Math. Phys.}  {\bf 44},  2149--2167 (2003) \href{https://doi.org/10.1063/1.1560552}{\path{doi:10.1063/1.1560552}}.


 
\bibitem{Kalnins1}
 Kalnins E.G.,  Pogosyan G.S.,  Miller W.Jr.:
  Completeness of multiseparable superintegrability on the complex 2-sphere.
{J. Phys. A: Math. Gen.} {\bf 33},  6791--6806 (2000) \href{https://doi.org/10.1088/0305-4470/33/38/310}{\path{doi:10.1088/0305-4470/33/38/310}}.


\bibitem{Kalnins2}
 Kalnins E.G.,    Kress J.M.,    Pogosyan G.S.,  Miller W.Jr.:  
  Completeness of superintegrability in two-dimensional
constant-curvature spaces.
{J. Phys. A: Math. Gen.} {\bf 34}, 4705--4720 (2001) \href{https://doi.org/10.1088/0305-4470/34/22/311}{\path{doi:10.1088/0305-4470/34/22/311}}.

 
\bibitem{kalnins}
Kalnins E.G., Benenti S., Miller W.Jr.: Integrability, {S}t\"ackel spaces, and rational potentials. {J. Math. Phys.} {\bf  38}, 2345--2365 (1997) \href{https://doi.org/10.1063/1.531977}{\path{doi:10.1063/1.531977}}.

  
\bibitem{Saksida}  Saksida  P.:   Integrable anharmonic oscillators on spheres and hyperbolic spaces.   {Nonlinearity} {\bf  14},  977--994 (2001) \href{https://doi.org/10.1088/0951-7715/14/5/304}{\path{doi:10.1088/0951-7715/14/5/304}}.


   \bibitem{Nersessian}
Nerssesian  A.,  Yeghikyan V.:   
Anisotropic inharmonic {H}iggs oscillator and related ({MICZ}-){K}epler-like systems. {J. Phys. A: Math. Theor.} {\bf 41},   155203 (2008) \href{https://doi.org/10.1088/1751-8113/41/15/155203}{\path{doi:10.1088/1751-8113/41/15/155203}}.

    

\bibitem{Marquette1}   Marquette I.:
Generalized MICZ-Kepler system, duality, polynomial, and deformed oscillator algebras. 
J. Math. Phys.   {\bf 51}, 102105  (2010)   \href{https://doi.org/10.1063/1.3496900}{\path{doi:10.1063/1.3496900}}.


\bibitem{Leemon}
 Leemon  H.I.: 
 Dynamical symmetries in a spherical geometry II.
{J. Phys.  A: Math. Gen.}  {\bf 12},  489--501 (1979) \href{https://doi.org/10.1088/0305-4470/12/4/009}{\path{doi:10.1088/0305-4470/12/4/009}}.



\bibitem{Pogoa}
Hakobyan Ye.M., Pogosyan   G.S., Sissakian  A.N.,  Vinitsky S.I.: Isotropic oscillator in a space of constant positive curvature: Interbasis expansions. {Phys. Atom. Nucl.} {\bf 62},   623--637  (1999) 
\newblock \href {https://arxiv.org/abs/quant-ph/9710045} {\path{arXiv:quant-ph/9710045}}.

  
 
\bibitem{Nersessian1}
 Nersessian A., Pogosyan G.:  Relation of the oscillator and Coulomb systems on spheres and pseudospheres.  {Phys. Rev. A} {\bf 63},    020103 (2001) \href{https://doi.org/10.1103/PhysRevA.63.020103}{\path{doi:10.1103/PhysRevA.63.020103}}.

\bibitem{Ranran}
Cari\~nena J.F.,    Ra\~nada  M.F.,    Santander M.,   Senthilvelan  M.: A non-linear oscillator with quasi-harmonic
behaviour: two- and n-dimensional oscillators.  {Nonlinearity}  {\bf 17},  1941-1963  (2004) \href{https://doi.org/10.1088/0951-7715/17/5/019}{\path{doi:10.1088/0951-7715/17/5/019}}.

 
  \bibitem{Annals09}
Ballesteros A., Enciso A., Herranz F.J., Ragnisco  O.:
Superintegrability on {N}-dimensional curved spaces: {C}entral potentials, centrifugal terms and monopoles. {Ann. Phys.}  {\bf 324},   1219--1233 (2009) \href{https://doi.org/10.1016/j.aop.2009.03.001}{\path{doi:10.1016/j.aop.2009.03.001}}.


\bibitem{Pogosyan1}
 Grosche C.,  Pogosyan G.S.,   Sissakian A.N.: 
  Path integral discussion for Smorodinsky--Winternitz  potentials
II. The two- and three-dimensional   sphere.
{Fortschr. Phys.} {\bf 43}  523--563 (1995)
 \href{https://doi.org/10.1002/prop.2190430603}{\path{doi:10.1002/prop.2190430603}}.
 
 
 
 
 \bibitem{HH}  H\'enon M.,  Heiles  C.: The applicability of the third integral of motion: Some numerical experiments. {Astron. J.} {\bf 69}, 73--79 (1964)
  \href{https://doi.org/10.1086/109234}{\path{doi:10.1086/109234}}.
  
  


   \bibitem{BSV} Bountis T.,  Segur H.,    Vivaldi F.:  Integrable {H}amiltonian systems and the {P}ainlev\'e property. {Phys. Rev.} A
\textbf{25},  1257--1264 (1982) \href{https://doi.org/10.1103/PhysRevA.25.1257}{\path{doi:10.1103/PhysRevA.25.1257}}.

   \bibitem{CTW}  Chang Y.F., Tabor M.,  Weiss J.:  Analytic structure of the {H}\'enon--{H}eiles Hamiltonian in integrable and nonintegrable regimes.
{J. Math. Phys.} \textbf{23},  531--538 (1982) \href{https://doi.org/10.1063/1.525389}{\path{doi:10.1063/1.525389}}.

   \bibitem{GDP} Grammaticos B.,  Dorizzi  B.,   Padjen R.:  {P}ainlev\'e property and integrals of motion for the {H}\'enon--{H}eiles system. {Phys. Lett.} A \textbf{89},   111--113 (1982) \href{https://doi.org/10.1016/0375-9601(82)90868-4}{\path{doi:10.1016/0375-9601(82)90868-4}}.

   \bibitem{HietarintaRapid}  Hietarinta  J.:  Integrable families of {H}\'enon--{H}eiles-type {H}amiltonians and a new duality.  {Phys. Rev.} A {\bf 28}   3670--3672 (1983) \href{https://doi.org/10.1103/PhysRevA.28.3670}{\path{doi:10.1103/PhysRevA.28.3670}}.

\bibitem{Fordy83} Fordy  A.P.:      
     {H}amiltonian symmetries of the {H}\'enon--{H}eiles system.
{Phys. Lett.} A \textbf{97}, 21--23  (1983) \href{https://doi.org/10.1016/0375-9601(83)90091-9}{\path{doi10.1016/0375-9601(83)90091-9}}.

\bibitem{Wojc}  Wojciechowski S.: Separability of an integrable case of the {H}\'enon--{H}eiles system.
{Phys. Lett.} A \textbf{100},  277--278  (1984) \href{https://doi.org/10.1016/0375-9601(84)90535-8}{\path{doi10.1016/0375-9601(84)90535-8}}.


\bibitem{SL}  Sahadevan R.,  Lakshmanan  M.:   Invariance and integrability: {H}\'enon--{H}eiles and two coupled quartic anharmonic oscillator systems.   {J. Phys. A: Math. Gen.} {\bf 19}, L949--L954 (1986)  \href{https://doi.org/10.1088/0305-4470/19/16/001}{\path{doi:10.1088/0305-4470/19/16/001}}.


\bibitem{FordyHH} Fordy A.P.:  The {H}\'enon--{H}eiles system revisited.
{Physica} D \textbf{52}, 204--210  (1991) \href{https://doi.org/10.1016/0167-2789(91)90122-P}{\path{doi:10.1016/0167-2789(91)90122-P}}.


\bibitem{Sarlet} Sarlet  W.:   New aspects of integrability of generalized {H}\'enon--{H}eiles systems. {J. Phys. A: Math. Gen.} {\bf 24},   5245--5251 (1991) \href{https://doi.org/10.1088/0305-4470/24/22/008}{\path{doi:10.1088/0305-4470/24/22/008}}.


\bibitem{RGG} Ravoson  V.,  Gavrilov L., Caboz R.:  Separability and Lax pairs for {H}\'enon--{H}eiles system.  {J. Math. Phys.}
\textbf{34}  2385--2393  (1993) \href{https://doi.org/10.1063/1.530123}{\path{doi:10.1063/1.530123}}.

 \bibitem{Tondo}  Tondo G.: On the integrability of stationary and restricted flows of the {K}d{V} hierarchy. 
 {J. Phys. A: Math. Gen.} \textbf{28},  5097--5115  (1995) \href{https://doi.org/10.1088/0305-4470/28/17/034}{\path{doi:10.1088/0305-4470/28/17/034}}.


\bibitem{Conte}  Conte  R., Musette  M.,   Verhoeven  C.:  Completeness of the cubic and quartic {H}\'enon--{H}eiles {H}amiltonians.  {Theor. Math. Phys.} \textbf{144},  888--898 (2005) \href{https://doi.org/10.1007/s11232-005-0115-9}{\path{doi:10.1007/s11232-005-0115-9}}.

  
   \bibitem{HHjpcs}
Ballesteros A., Blasco A., Herranz F.J.:
  A curved H\'enon-Heiles system and its integrable perturbations
{J. Phys.: Conf. Ser.} {\bf 597}  012013 (2015)
 \href{https://doi.org/10.1088/1742-6596/597/1/012013}{\path{doi:10.1088/1742-6596/597/1/012013}}.
 



 \bibitem{RDGprl}
Ramani A.,  Dorizzi  B.,   Grammaticos B.:   {P}ainlev\'e conjecture revisited. {Phys. Rev. Lett.}  {\bf 49},  1539--1541 (1982) \href{https://doi.org/10.1103/PhysRevLett.49.1539}{\path{doi:10.1103/PhysRevLett.49.1539}}.

\bibitem{Hietarinta}  Hietarinta  J.:   Direct method for the search of the second invariant.   {Phys. Rep.} \textbf{147},  87--154 (1987) \href{https://doi.org/10.1016/0370-1573(87)90089-5}{\path{doi:10.1016/0370-1573(87)90089-5}}.

   
    \bibitem{FF}
 Ferapontov  E.V.,  Fordy  A.P.:   Separable {H}amiltonians and integrable systems of hydrodynamic type.   {J. Geom. Phys.}   {\bf 21},   169--182 (1997) \href{https://doi.org/10.1016/S0393-0440(96)00013-7}{\path{doi:10.1016/S0393-0440(96)00013-7}}.

\bibitem{HoneIP}
 Hone  A.N.W.,  Novikov  V.,   Verhoeven C.: An integrable hierarchy with a perturbed {H}\'enon--{H}eiles system.  {Inv. Probl.} {\bf 22}, 
 2001--2020 (2006)  \href{https://doi.org/10.1088/0266-5611/22/6/006}{\path{doi:10.1088/0266-5611/22/6/006}}.

\bibitem{HonePLA}
 Hone   A.N.W., Novikov V.,  Verhoeven C.:  An extended {H}\'enon--{H}eiles system.   {Phys. Lett.}  A {\bf 372},   1440--1444 (2008) \href{https://doi.org/10.1016/j.physleta.2007.09.063}{\path{doi:10.1016/j.physleta.2007.09.063}}.



\bibitem{tesis}
 Blasco A.:   \textit{Integrability of non-linear {H}amiltonian systems with $N$ degrees of freedom}  PhD Thesis (Burgos: Burgos University) (2009) \href{http://riubu.ubu.es/bitstream/10259/106/4/Blasco_Sanz.pdf}{\path{http://riubu.ubu.es/bitstream/10259/106/4/Blasco_Sanz.pdf}}.


\bibitem{Annals10} Ballesteros  A.,  Blasco  A.:  Integrable {H}\'enon--{H}eiles {H}amiltonians: a {P}oisson algebra approach. {Ann.
Phys.} {\bf 325}, 2787--2799 (2010) \href{https://doi.org/10.1016/j.aop.2010.08.002}{\path{doi:10.1016/j.aop.2010.08.002}}.



  \bibitem{HHnon}   Ballesteros A.,  Blasco  A., Herranz   F.J., Musso  F.: An integrable H\'enon--Heiles system on the sphere and the hyperbolic plane.  {Nonlinearity} \textbf{28},  3789--3801 (2015)    
 \href{https://doi.org/10.1088/0951-7715/28/11/3789}{\path{doi:10.1088/0951-7715/28/11/3789}}.


 
   \bibitem{Serret} Serret P.: Th\'eorie nouvelle g\'eom\'etrique et m\'ecanique des lignes \`a double courbure. Paris, Mallet-Bachelier (1859).
   
       
 \bibitem{Petrosian1}   Petrosyan D.R.,    Pogosyan  G.S.: 
Harmonic oscillator on the SO(2,2) hyperboloid. 
 SIGMA  Symmetry Integrability Geom. Methods Appl. {\bf  11},  096  (2015) \href{https://doi.org/10.3842/SIGMA.2015.096}{\path{doi:10.3842/SIGMA.2015.096}}.

     
   
\bibitem{KaKrWint02}   Kalnins  E.G.,    Kress J.M.,    Winternitz  P.:
Superintegrability in a two-dimensional space of nonconstant curvature. 
J. Math. Phys. {\bf  43},     970--983  (2002) \href{https://doi.org/10.1063/1.1429322}{\path{doi:10.1063/1.1429322}} .
     
\bibitem{intQG}  Ballesteros A.,    Herranz F.J.,    Ragnisco  O.: 
 Integrable potentials on spaces with curvature from quantum groups. 
  {J. Phys. A: Math. Theor.} \textbf{38},  7129--7144 (2005)    
 \href{https://doi.org/10.1088/0305-4470/38/32/004}{\path{doi:10.1088/0305-4470/38/32/004}}.
     
\bibitem{PhysD2008}     Ballesteros A.,  Enciso A.,    Herranz  F.J.,    Ragnisco O.: 
A maximally superintegrable system on an n-dimensional space of nonconstant curvature.
Physica D {\bf 237},   505--509  (2008) \href{https://doi.org/10.1016/j.physd.2007.09.021}{\path{doi:10.1016/j.physd.2007.09.021}}.

\bibitem{Ballestetal08ClQGr}   Ballesteros A.,   Enciso  A.,      Herranz  F.J.,     Ragnisco   O.:
{B}ertrand spacetimes as {K}epler/oscillator potentials. 
Class. Quantum Grav. {\bf  25},   165005  (2008) \href{https://doi.org/10.1088/0264-9381/25/16/165005}{\path{doi:10.1088/0264-9381/25/16/165005}}.    

\bibitem{ComRunge}   Ballesteros  A.,   Enciso A.,      Herranz F.J.,   Ragnisco   O.:
{H}amiltonian systems admitting a {R}unge-{L}enz vector and an optimal extension of {B}ertrand's theorem to curved manifolds.    
 Comm. Math. Phys.  {\bf  290},    1033--1049  (2009) \href{https://doi.org/10.1007/s00220-009-0793-5}
 {\path{doi:10.1007/s00220-009-0793-5}}
 
 
 
 \bibitem{WHS}   Ballesteros  A.,   Blasco A.,      Herranz F.J.,  Musso F.,  Ragnisco   O.:
(Super)integrability from coalgebra symmetry: formalism and applications.    
 J.  Phys.: Conf. Ser.  {\bf  175},   012004  (2009) \href{https://doi.org/10.1088/1742-6596/175/1/012004}
 {\path{doi:10.1088/1742-6596/175/1/012004}}

 
 
  \bibitem{RagRig10}    Ragnisco O.,   Riglioni  D.:
A family of exactly solvable radial quantum systems on space of non-constant curvature with accidental degeneracy in the spectrum.   
 {SIGMA} Symmetry Integrability Geom. Methods Appl. {\bf 6},   097  (2010) \href{https://doi.org/10.3842/SIGMA.2010.097}{\path{doi:10.3842/SIGMA.2010.097}}.      
  
  \bibitem{Sigma11}   Ballesteros A.,    Enciso  A.,      Herranz F.J.,    Ragnisco O.,    Riglioni O.: 
Superintegrable oscillator and {K}epler systems on spaces of nonconstant curvature via the {S}\"ackel Transform. 
 SIGMA  Symmetry Integrability Geom. Methods Appl. {\bf  7},  048  (2011) \href{https://doi.org/10.3842/SIGMA.2011.048}{\path{doi:10.3842/SIGMA.2011.048}}.

\bibitem{BurgosAnnPh11}   Ballesteros  A.,    Enciso  A.,    Herranz  F.J.,    Ragnisco  O.,    Riglioni  O.:
Quantum mechanics on spaces of nonconstant curvature: the oscillator problem and superintegrability. 
 Ann. Phys.  {\bf  326},     2053--2073  (2011) \href{https://doi.org/10.1016/j.aop.2011.03.002}{\path{doi:10.1016/j.aop.2011.03.002}}.

  \bibitem{taub}
  Ballesteros  A., Enciso  A., Herranz  F.J., Ragnisco  O., Riglioni  D.: An exactly solvable deformation of the {C}oulomb problem associated with the {T}aub-{NUT} metric.
{Ann. Phys.}  {\bf 351},  540--577 (2014) \href{https://doi.org/10.1016/j.aop.2014.09.013}{\path{doi:10.1016/j.aop.2014.09.013}}.
 
 
 
 %%%%%%%%%%%%%%%%%%%%%%%%%%%%%%%%%%%%%%%%%%%




\end{thebibliography}
\end{document}